\newif\ifarxiv
\arxivtrue   

\ifarxiv
  \documentclass[journal,onecolumn,11pt]{IEEEtran}
\else
  \documentclass[lettersize,journal]{IEEEtran}
\fi

\usepackage{cmap}
\usepackage[T1]{fontenc}
\usepackage{amsmath,amsfonts,amssymb}
\usepackage{array}
\usepackage{booktabs}
\usepackage{textcomp}
\usepackage{stfloats}
\usepackage{graphicx}
\usepackage{cite}
\usepackage{orcidlink}
\usepackage{balance}
\PassOptionsToPackage{hyphens}{url}
\usepackage{url}
\usepackage{xurl}

\ifarxiv
  \newcommand{\figcolw}{4.4in}
  \newcommand{\figwidew}{\linewidth}
\else
  \newcommand{\figcolw}{3.4in}
  \newcommand{\figwidew}{6.9in}
\fi

\begin{document}

\ifarxiv \sloppy \fi

\title{Mode~0: Architecture, Risk Taxonomy, and Standardization Pathway
       for RCU-Assisted V2X Safety Communication}

\author{Dewei~Jiang\,\orcidlink{0009-0000-3698-0248}\
        and~Xiang~Gu\,\orcidlink{0000-0002-2594-4118}
\thanks{Manuscript prepared October 2026.
        \emph{(Corresponding author: Xiang Gu.)}}%
\thanks{Dewei Jiang is with the School of Information Science and
        Technology, Nantong University, Nantong, China
        (e-mail: 6302021207@stmail.ntu.edu.cn).}%
\thanks{Xiang Gu is with the School of Artificial Intelligence and
        Computer Science, Nantong University, Nantong, China
        (e-mail: gu.x@ntu.edu.cn).}}

\ifarxiv
  \markboth{Mode~0: Architecture, Risk Taxonomy, and Standardization
            Pathway for RCU-Assisted V2X Safety Communication}%
           {Jiang and Gu: Mode~0 --- Architecture, Risk Taxonomy, and
            Standardization Pathway}
\else
  \markboth{Journal Manuscript, Vol.~XX, No.~Y, Month 20XX}%
           {Jiang and Gu: Mode~0 --- Architecture, Risk Taxonomy, and
            Standardization Pathway}
\fi

\maketitle

\begin{abstract}
The 3GPP V2X resource allocation framework offers two entity classes for
coordinating vehicle communication --- base-station scheduling and
autonomous vehicle selection --- a design space we show to be structurally
incomplete for safety-critical coordination. We propose Mode 0, a third
entity class centered on the Roadside Computing Unit (RCU): infrastructure
integrating elevated sensing, sidelink communication, and local
computation, owned by traffic management authorities rather than network
operators. We formalize Mode 0's subfamily taxonomy (Mode 0a through 0c),
its traffic risk taxonomy and mandatory-escalation boundary, its
cross-domain coexistence with Modes 1--4, and its standardization pathway.
Of the three failure modes motivating this proposal --- scheduling
saturation, occlusion-driven information gaps, and an institutional
authority gap during severe hazards --- only scheduling is evaluated by
simulation; information and authority are argued architecturally and
evidenced by independent deployment convergence across Chinese local and
industry-alliance standards, Chinese operator infrastructure, and European
and US C-V2X programs, properties no channel-level simulation can
instantiate. In a multi-agent reinforcement learning simulation (88
training runs, 95\% confidence intervals), per-vehicle policies (Mode 0c)
raise worst-interval delivery reliability at the critical safety load from
0.217 under shared per-class policies (Mode 0a) to 0.792, cutting lost
safety messages from 38\% to 6\%. An idealized model of standardized
sensing-based semi-persistent scheduling loses 10\% at the same load, and
per-vehicle coordination more than halves the longest gaps between safety
updates against it (3.9 against 8.4 message periods). With spare spectrum,
a safety partition sized by the RCU delivers every safety message. We call
for a 3GPP study item to formalize Mode 0.
\end{abstract}


\section{Introduction}\label{sec:1}

Every second of warning time before a vehicle collision corresponds to approximately 30 meters of travel at highway speeds. The difference between a near-miss and a fatality is often measured not in seconds but in fractions of seconds --- which is why 3GPP safety specifications bound the end-to-end latency of pre-crash sensing warnings at 20 milliseconds~\cite{ref1}. This is not a conservative engineering margin. It is a threshold derived directly from the physics of human reaction time and vehicle dynamics, below which warning delivery is useful and above which it may arrive too late to prevent harm. Any V2X communication architecture that cannot guarantee this bound as a worst-case property --- not merely as an average --- is architecturally insufficient for the safety applications it is intended to serve.

The 3GPP V2X resource allocation framework provides two architectural patterns for meeting this requirement for safety traffic, which includes periodic awareness messages at 1--10 Hz per vehicle~\cite{ref2} as well as event-driven hazard warnings. In Mode 3 (Long Term Evolution V2X, LTE-V2X) and Mode 1 (New Radio V2X, NR-V2X), the base station --- the evolved Node B (eNB) and the next-generation Node B (gNB), respectively --- exercises centralized scheduling authority over all vehicle transmissions within its coverage area. In Mode 4 (LTE-V2X) and Mode 2 (NR-V2X), vehicle user equipment (UE) exercise autonomous distributed resource selection without infrastructure involvement (Section~\ref{sec:3.1}). Despite the evolution from LTE to NR and the introduction of increasingly sophisticated autonomous selection mechanisms, these two patterns exhaust the design space as currently standardized: every defined mode reduces to centralized base station authority or distributed UE autonomy. This paper demonstrates that both patterns fail under the density and information conditions that define the most safety-critical V2X scenarios, and that the failure is structural rather than parametric --- it cannot be resolved by optimizing the schedulers within either existing entity class. The entity capable of resolving it belongs to a third class that has no standards home in the current framework.

\begin{figure}[!t]
\centering
\includegraphics[width=\figcolw,keepaspectratio]%
                {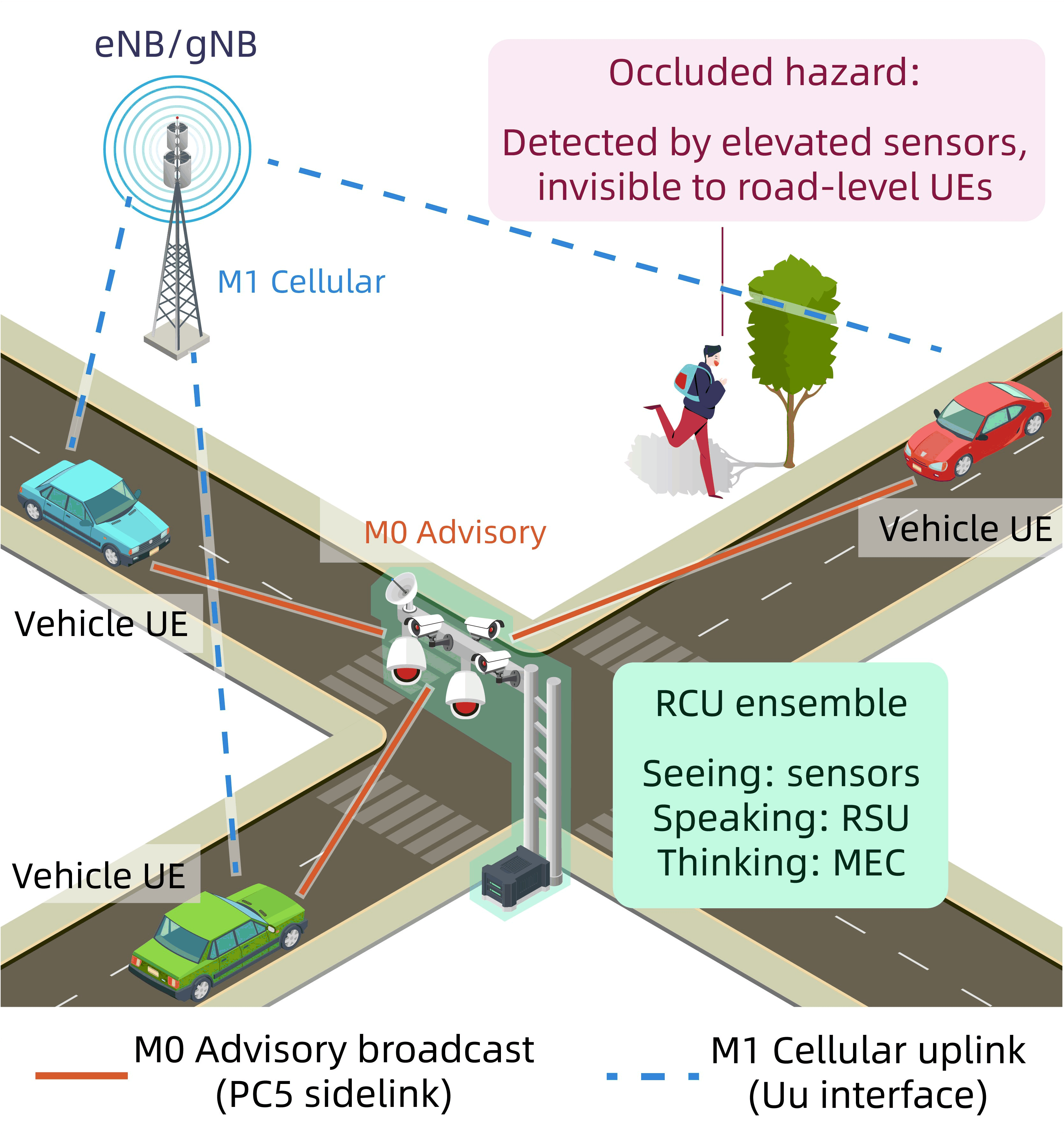}
\caption{Mode 0 operational scenario at a traffic intersection. The Roadside Computing Unit (RCU) --- an infrastructure ensemble integrating elevated sensing (Seeing), sidelink communication (Speaking), and local computation (Thinking) --- broadcasts M0 safety-critical advisory messages to vehicle UEs. In this illustration, the RCU's messages use PC5 sidelink (solid red lines) and M1 non-safety traffic uses cellular uplinks (Uu interface) to the eNB/gNB (dashed blue lines); the M0/M1 classification itself does not depend on the interface (Section~\ref{sec:4.2}). The RCU's observation layer detects an occluded pedestrian (a ghost peeking hazard: a pedestrian stepping out from behind an obstruction) invisible to road-level vehicle sensors, providing pre-emergence hazard warning capability structurally inaccessible to Mode 4/2 UE autonomy.}
\label{fig:1}
\end{figure}

We propose Mode 0, a new 3GPP V2X resource allocation category whose defining entity is the Roadside Computing Unit (RCU). The RCU is not merely a roadside unit (RSU) with added compute, and the RSU-MEC or edge-RSU compound used in the literature~\cite{ref3,ref4} --- an RSU paired with a multi-access edge computing (MEC) host --- does not describe it: that compound specifies a communication layer and an evaluation layer but omits the observation layer. As Section~\ref{sec:5} argues, the RCU is an infrastructure ensemble of three functionally necessary components: an observation layer of elevated sensors observes the zone (its modalities are a deployment choice rather than an architectural requirement); a communication layer, instantiated by a roadside unit at the traffic node, broadcasts advisory messages to vehicle UEs, aggregates V2X inputs from the zone, and exchanges coordination data with the network operator and traffic management platforms; and an evaluation layer, instantiated by a MEC facility, fuses sensor and V2X data into a zone model and derives sub-zone resource recommendations. The observation layer is what makes the RCU's most important capabilities possible: a regional mobility model, and pre-emergence detection of hazards to occluded traffic participants. The RCU is fixed traffic infrastructure, owned and operated by a traffic management authority. It is neither a base station nor a UE; its objective is traffic safety and flow rather than network throughput; and its regulatory home is transportation law rather than telecommunications law. Mode 0 is advisory by default, like every existing 3GPP mode, with an emergency escalation protocol for large-scope hazards (Sections~\ref{sec:4.1} and~\ref{sec:5.2}). It is complementary to Modes 1 through 4 and independent of any particular radio generation. Fig.~\ref{fig:1} illustrates the operational scenario.

Prior work on V2X resource allocation optimizes within the two entity classes that 3GPP defines --- base-station scheduling and autonomous UE selection. This paper's contribution is at the level of that taxonomy. It argues that the two classes are structurally incomplete for safety-critical coordination along three dimensions --- scheduling saturation, occlusion-driven information gaps, and the authority gap in large-scope hazards --- and proposes a third entity class, the traffic-authority-owned RCU, together with its M0/M1 information classification and demand separation architecture, its subfamily taxonomy, its traffic risk taxonomy and mandatory-escalation boundary, its coexistence with Modes 1--4, and a standardization pathway. Of the three dimensions, scheduling is evaluated by simulation against the standardized baseline; information and authority are argued architecturally and supported by deployment evidence.

This positioning is distinct from three adjacent bodies of work, detailed in Section~\ref{sec:2}. Existing RSU-MEC and edge-RSU proposals~\cite{ref3,ref4} use roadside compute to improve allocation within the existing base-station and UE-autonomous categories, and omit the observation layer on which the RCU's pre-emergence hazard detection and escalation authority depend. Reinforcement-learning approaches to NR Mode 2~\cite{ref5} refine UE-autonomous selection within its information constraints, which are categorical rather than algorithmic: no learned policy gives a mobile terminal the elevated vantage point that an occluded hazard requires. The geographic resource zones of LTE-V2X Release 14~\cite{ref6} divide space but not traffic class, whereas Mode 0 provisions pools by safety class, adjusts their granularity to the zone, and issues recommendations at every interval. A reading of the survey in~\cite{ref5}, which covers over 190 works, indicates that none of them addresses a traffic-authority-owned entity class, a safety-class pool partition at the infrastructure level, or an emergency escalation protocol (Section~\ref{sec:2}).

The simulation evaluates the scheduling dimension with Multi-Agent Proximal Policy Optimization (MAPPO) under a Centralized Training Distributed Execution (CTDE) architecture, with capability ablations and confidence intervals over independent seeds. Its baseline is an idealized model of sensing-based semi-persistent scheduling (SB-SPS), the NR-V2X Mode 2 mechanism; Section~\ref{sec:6.7} lists the idealizations, which favor SB-SPS in some respects and Mode 0 in others and on balance flatter Mode 0. The simulation establishes four findings. (i) Per-vehicle policies become necessary once a class group is too large for a single shared policy to separate; this happens at and above the critical safety load, where the number of safety vehicles equals the number of subchannels. (ii) Against SB-SPS, the margin of per-vehicle policies is largest in the tail --- at the worst interval, for the worst vehicle, and in peak information age --- and it comes from coordinated subchannel choice rather than power control. (iii) The measured contribution of the RCU lies in execution: a safety partition sized to the safety demand, imposed only where spectrum is spare, while the global view of the critic brings no measurable advantage in training. (iv) Mixed fleets keep safety delivery for passive and active vehicles alike.

Scheduling, information, and authority admit different kinds of evidence, and this paper is explicit about which is which. Scheduling saturation is evaluated by the proof-of-concept simulation described above, under a declared scope: a single highway zone with one contention pool and a compressed resource grid. How a traffic authority sizes and distributes a safety partition in practice is addressed in Section~\ref{sec:8.9}. The information gap --- the sensing and geometric advantage that an elevated, zone-wide vantage point holds over any mobile terminal --- is a property of physical sensing infrastructure, and the authority gap --- the institutional mandate required to escalate from advisory to mandatory control during large-scope hazards --- is a property of institutions. No channel-level simulation can instantiate either, so both are argued architecturally (Sections~\ref{sec:3}--\ref{sec:5}) and grounded in deployment evidence. Chinese local and industry-alliance standards, including DB11/T 2329.1-2024~\cite{ref7} (a Beijing municipal standard) and T/ITS 0224.1-2025~\cite{ref8} (an industry-alliance standard), and European Cooperative Intelligent Transport Systems (C-ITS) and United States cellular V2X (C-V2X) infrastructure programs show multiple independent regulatory ecosystems converging on an entity matching the RCU specification, deployed at scale without a 3GPP mode to inhabit. Network operators such as China Unicom have meanwhile deployed operator-owned roadside multi-access edge computing units (RS-MEC~\cite{ref9}) at the same locations, so both sides of the institutional boundary are present at the traffic node and a standardized coordination mechanism is needed. Mode 0 standardization would give these deployments a framework for interoperability, security specification, and coexistence with 3GPP vehicle communication.

The remainder of this paper is organized as follows. Section~\ref{sec:2} reviews related work. Section~\ref{sec:3} analyzes the structural failure modes of the existing 3GPP V2X mode taxonomy along these three dimensions. Section~\ref{sec:4} makes the affirmative case for Mode 0 through the compute sinking argument, the M0/M1 information classification and demand separation imperative, and the convergent deployment evidence. Section~\ref{sec:5} defines Mode 0 and the RCU entity class, including the emergency escalation protocol, the compliance model, coexistence with existing modes, and data locality. Section~\ref{sec:6} presents the system model and the simulation design, and Section~\ref{sec:7} reports seven results, from the critical safety load to information age. Section~\ref{sec:8} addresses the standardization pathway, the cross-domain interface, compliance incentives, the generational roadmap, the vehicle-to-pedestrian (V2P) extension, the Mode 0d voluntary downgrade question, mobile deployment, security of the escalation channel, pool sizing, and open issues. Section~\ref{sec:9} concludes.

\section{Related Work}\label{sec:2}

\textbf{Resource allocation in 3GPP V2X networks.} The 3GPP V2X resource allocation framework has been the subject of extensive study across both its generational instantiations. Nair and Tanwar~\cite{ref5} provide a comprehensive taxonomy --- covering graph-based, game-theoretic, genetic algorithm, optimization, machine learning, deep learning, and reinforcement learning approaches to power, channel, and spectrum allocation --- that confirms the binary structure of the current framework: network-scheduled modes (LTE Mode 3 / NR Mode 1) and UE-autonomous modes (LTE Mode 4 / NR Mode 2). Their survey identifies interference and resource collision under UE-autonomous distributed selection as the dominant open challenge, a finding echoed in RL-based approaches that address NR Mode 2 sensing-based semi-persistent scheduling inaccuracies under dense and aperiodic traffic~\cite{ref5}. These works motivate the structural gap analysis of Section~\ref{sec:3} but do not propose a third entity class or a safety-class resource isolation mechanism to resolve it.

\textbf{Sensing-based scheduling and learned Mode 2 selection.} In NR-V2X Mode 2, a vehicle selects its own sidelink resources using either sensing-based semi-persistent scheduling (SB-SPS) or dynamic scheduling, and the specifications do not state when each should be used~\cite{ref10}. SB-SPS reserves resources for periodic traffic, so it suits the periodic, fixed-size safety messages modeled in Section~\ref{sec:6.3}, but it performs poorly under aperiodic traffic with variable message sizes. Dynamic scheduling copes better with such traffic but, because it selects resources at random, suffers more collisions when density is high~\cite{ref11}. Learned schemes have been proposed to reduce these collisions; one example is a decentralized multi-agent actor-critic method, reported to outperform both SB-SPS and dynamic scheduling under aperiodic traffic~\cite{ref11}. The MARL approaches discussed below take the same stance: the decision is made by the vehicles, from information gathered at road level, a limit that Section~\ref{sec:3.3} argues cannot be removed by better algorithms, and to our knowledge none of these schemes uses an elevated, zone-wide source of information. What Mode 0 adds to learned scheduling is the RCU's zone-level execution functions, not a better learning algorithm. We therefore use SB-SPS, with idealized sensing (Section~\ref{sec:6.7}), as the standardized reference point of Section~\ref{sec:7}, and the critic-input ablation of Section~\ref{sec:7.5} tests what global information adds beyond fully local learning.

\textbf{Zone-based resource pools in LTE-V2X.} LTE-V2X introduced geographic resource zones in Release 14, in which a vehicle selects its transmission pool according to the zone it occupies~\cite{ref6}. A reader could therefore take a Mode 0 safety partition for a Release 14 feature. The two differ in four respects: Mode 0 pools are defined by traffic class rather than by geography alone; their granularity adapts to the traffic node; the RCU issues recommendations at every generation interval rather than relying on a static configuration; and the RCU recommends placement within a pool from the positions of vehicles it observes.

\textbf{RSU-assisted and infrastructure-coordinated approaches.} Prior work has explored roadside infrastructure as a coordination mechanism for vehicular resource allocation. RSUs equipped with cognitive radio capabilities have been applied to spectrum management in heterogeneous vehicular networks through game-theoretic non-cooperative approaches that optimize aggregate throughput across vehicles served by different RSUs~\cite{ref5}. Infrastructure-assisted cooperative perception through RSU relay and data fusion has been studied as a means to extend vehicles' effective sensor range beyond direct line-of-sight limitations~\cite{ref3}. These proposals share surface characteristics with Mode 0 --- fixed roadside coordination points, elevated information availability --- but differ in three critical respects: they are operated by network operators rather than traffic management authorities; they optimize aggregate throughput rather than safety-class delivery guarantees; and they lack the formalized third entity class, the M0/M1 resource pool partition, and the mandatory emergency escalation protocol that Mode 0 introduces. No prior work, to the authors' knowledge, proposes a traffic-authority-owned, sensing-equipped roadside entity as a new 3GPP V2X resource allocation category.

\textbf{Multi-agent reinforcement learning for V2X resource allocation.} MARL has emerged as a promising framework for distributed V2X resource coordination. Vu et al.~\cite{ref12} apply MARL to channel assignment and power allocation in platoon-based C-V2X systems, with each platoon member acting as an independent learning agent, and Cui et al.~\cite{ref13} apply it to the joint allocation of spectrum and computing resources by base stations in vehicular networks. Attention-mechanism MARL architectures have been proposed to enable inter-agent focus on relevant neighbor states for joint spectrum and power selection, and MARL combined with mean-field game theory has been shown to reduce computational complexity for large vehicle populations while approaching near-optimal performance~\cite{ref5}. The vehicle-side approaches operate under fully decentralized execution consistent with NR-V2X Mode 2 UE-autonomous operation, and do not incorporate a centralized information entity at execution time. Our MAPPO implementation employs a centralized critic during training, which is absent during execution~\cite{ref14}, and in Mode 0a it shares one actor among the vehicles of a class~\cite{ref15}. We treat the critic as a training device rather than as a model of the RCU, and Section~\ref{sec:7.5} tests directly whether its global information matters. The RCU enters the simulation through its execution-side functions: how vehicles are grouped under shared policies, and how a safety partition is sized (Sections~\ref{sec:7.6} and~\ref{sec:7.7}).

\textbf{Integrated sensing and communication in V2X.} Bartoletti et al.~\cite{ref16} analyze the use of NR Mode 2 sidelink signals for integrated sensing and communication (ISAC), demonstrating that the UE-autonomous distributed resource allocation mechanism produces interference that substantially degrades radar sensing performance relative to the interference-free bound, with vehicle density identified as the primary factor. This result is relevant to Mode 0 in two ways. First, it provides independent evidence that UE-autonomous allocation produces interference patterns that impair the broader information environment --- a finding complementary to our packet delivery tail results. Second, it establishes a distinction between vehicle-embedded ISAC and infrastructure sensing: the Mode 0 RCU's observation layer uses dedicated infrastructure sensors at intersection height rather than repurposing communication signals, providing an elevation advantage and independence from the co-channel interference environment that vehicle-mounted ISAC cannot achieve.

\textbf{Positioning of this work.} The survey by Nair and Tanwar~\cite{ref5} covers over 190 resource allocation papers across all major algorithmic families and V2X application scenarios. None of the covered approaches addresses: (i) a traffic-authority-owned, sensing-equipped infrastructure entity as a formal 3GPP entity class; (ii) an M0/M1 safety-vs-non-safety resource pool partition implemented at the infrastructure level, decoupled from network operator optimization objectives; (iii) a two-pathway mandatory emergency escalation protocol for large-scope cascading hazards; or (iv) an evaluation that measures where shared-policy learning fails and what an infrastructure-sized safety partition adds. Mode 0's novelty is at the entity classification level --- a layer above the choice of resource allocation algorithm --- and is positioned as an architectural proposal with proof-of-concept simulation rather than as an algorithmic improvement within the existing entity taxonomy.

\section{Limitations of the Existing 3GPP V2X Resource Allocation Framework}\label{sec:3}

\subsection{The Existing Mode Taxonomy}\label{sec:3.1}

The 3GPP V2X resource allocation framework has evolved across two generations of cellular technology, producing four distinct operational modes that together define the complete space of standardized V2X communication architectures. Understanding the structural logic of this taxonomy is a prerequisite for identifying where it is incomplete.

In the LTE-V2X generation, Mode 3 places scheduling authority entirely within the evolved Node B (eNB). Vehicles transmit safety messages --- including Basic Safety Messages (BSMs) and Cooperative Awareness Messages (CAMs) --- only when the eNB grants the necessary uplink resources. The vehicle has no autonomous scheduling capability in this mode; its transmissions are gated by infrastructure. Mode 4 inverts this relationship entirely. Vehicles select their own transmission resources from a pre-configured pool without any infrastructure involvement, using sensing-based mechanisms to estimate channel occupancy and avoid collisions. The eNB plays no role in Mode 4 scheduling decisions.

The NR-V2X generation replicated this binary structure at higher capability~\cite{ref17}. Mode 1 assigns resource scheduling to the gNB, directly paralleling LTE Mode 3 with the performance advantages of New Radio. Mode 2 extends LTE Mode 4's autonomous selection principle to the NR air interface, incorporating an enhanced sensing-based resource selection mechanism designed to reduce inter-vehicle interference through more sophisticated channel monitoring.

Despite the generational progression from LTE to NR, and the sub-modes 2(a) to 2(d) of NR-V2X Mode 2 that the Release 16 study considered~\cite{ref18}, which vary UE-side behavior in different directions, the structural logic of the taxonomy has remained invariant. Every defined mode reduces to one of exactly two architectural patterns: a base station entity exercises centralized scheduling authority over vehicle UEs, or vehicle UEs exercise autonomous distributed scheduling without infrastructure involvement. These two patterns exhaust the design space as currently standardized. As we demonstrate in the following subsections, this binary taxonomy is insufficient for the safety-critical V2X scenarios that define the most demanding operational conditions, and a third architectural pattern --- one that has no standards home in the current framework --- is both necessary and increasingly present in deployed infrastructure.

\subsection{Resource Contention and the Structural Failure of Base Station-Led Scheduling}\label{sec:3.2}

The appeal of Mode 3/1 scheduling is straightforward: the base station possesses global visibility over all connected UEs within its cell, enabling interference-aware resource allocation that autonomous UEs cannot perform. For general cellular services, this centralized authority is architecturally appropriate. For safety-critical V2X communications at traffic nodes, it introduces a structural failure mode that cannot be resolved through parameter optimization alone.

The failure originates in resource contention. Under 3GPP specifications, safety applications including cooperative awareness require transmission frequencies of 1--10 Hz per vehicle, as specified in ETSI EN 302 637-2~\cite{ref2}. End-to-end latency requirements for cooperative awareness are bounded at 100 ms, with pre-crash sensing warnings requiring 20 ms or below, as established in 3GPP TR 22.885~\cite{ref1}. These requirements are demanding but individually manageable --- until the geometry of traffic nodes is considered.

At a signalized urban intersection, vehicle density within a 300-meter radius routinely exceeds 100 vehicles during peak hours. Each vehicle operating in Mode 3/1 must first transmit a Scheduling Request (SR) to the gNB, await a scheduling grant, and then transmit its safety message. Under normal cellular load, this process completes within 4--8 milliseconds. However, if 100 vehicles each need a dynamically scheduled grant at 10 Hz, the gNB faces a scheduling request load of 1,000 grant cycles per second from vehicular traffic alone --- before accounting for the pedestrian and general cellular traffic that inevitably co-occupies the same coverage zone. Periodic CAMs need not generate this load, because Mode 3 semi-persistent scheduling and Mode 1 configured grants serve them without a per-message request. The figure is therefore the load of dynamic grants, which event-driven warnings require because they cannot be scheduled in advance, and the argument that follows rests on that load and on the 20 ms bound those warnings must meet. Cellular capacity is provisioned for average conditions rather than peak ones, since spectrum and infrastructure expansion is costly and slow relative to demand growth~\cite{ref19}. The physical layer capacity of a single gNB, typically serving 500--1,000 active UEs depending on configuration, is exactly the kind of shared resource this dynamic strains: it approaches saturation precisely under peak conditions such as those described above. When contention-driven queuing delay is added to the inherent processing latency of the scheduling cycle, the 20 ms pre-crash requirement becomes difficult to guarantee as a worst-case bound rather than an average.

Critically, the co-presence of non-vehicular traffic is not an edge case but a systematic characteristic of precisely the environments where V2X safety requirements are most stringent. High-density traffic nodes --- intersections, merge points, transit hubs --- are embedded in urban environments where pedestrian density and general cellular traffic demand are simultaneously elevated. The 5G Quality of Service (QoS) Identifier (5QI) priority framework provides QoS differentiation within the gNB scheduler, ensuring that safety-class traffic is served before entertainment and other non-safety traffic within the available resource pool. However, priority scheduling cannot create scheduling capacity that does not physically exist. This distinction matters most precisely when it is hardest to act on. Priority determines which request is served the moment a resource becomes free; it has nothing to arbitrate when no resource is free at all. High-density conditions are precisely the conditions under which the latter case dominates: resources are already fully occupied, M0 and M1 requests queue simultaneously, and priority sits idle alongside them, with no available slot to sort. Each generation interval that passes without a resource freeing up does not merely delay the existing backlog --- it compounds it, since new M0 and M1 messages continue arriving every generation interval regardless of whether the previous ones were served. The queue that priority will eventually sort, once a resource does free up, has already grown larger in the interval.

When the air interface approaches saturation under the aggregate demand from safety messages, social media uploads, video streaming, and all other traffic types sharing the same resource pool on the same hardware, even correctly prioritized safety traffic experiences latency penalties that the 5QI framework cannot eliminate. The architectural problem is not the absence of priority mechanisms --- it is that M0 safety-critical traffic and M1 entertainment and non-safety traffic --- including navigation updates, over-the-air software downloads, and general cellular data --- share the same physical resources under the same scheduler at the same base station. This is a structural coupling that QoS policies can moderate but cannot sever.

The misalignment between the gNB's optimization objective and the V2X safety requirement further compounds this structural problem, though it is secondary to contention as a root cause. The gNB optimizes for aggregate performance across its full subscriber base --- a population whose traffic demands are dominated by M1-class services. Safety message scheduling is a minority use case in this optimization context. Adaptive scheduling algorithms and machine learning-based resource management can improve the prioritization of safety traffic, but these approaches address the symptom rather than the cause. They delay the onset of contention-driven latency degradation under increasing load; they do not eliminate the shared-resource competition that produces it.

\subsection{UE Autonomy, Global Blindness, and the Categorical Gap}\label{sec:3.3}

The Mode 4/2 architecture avoids the base station contention problem by removing the base station from the scheduling loop entirely. Vehicles assess channel occupancy through sensing, select resources from a pre-configured pool, and transmit without waiting for infrastructure grants. The NR-V2X Mode 2 sensing-based resource selection mechanism represents a mature engineering solution to the distributed coordination problem, and under moderate density conditions it performs well. Its limitations, however, are not incidental imperfections amenable to further refinement --- they are categorical consequences of the peer entity model on which NR-V2X Mode 2 is constructed.

The 3GPP standardization record is explicit on this point. The NR-V2X Mode 2 family, including its most behaviorally sophisticated sub-mode, Mode 2(d), which the Release 16 study considered~\cite{ref18}, is built on the principle that all participating UEs are entities of the same class. Document R1-1812399, which contributed to the design of NR-V2X Mode 2 (UE-autonomous) resource allocation, states this directly: ``The group leader UE is by no means a different class or category of UEs, as all UEs are expected to have the capability to be a group leader''~\cite{ref20}. The Mode 2(d) group leader has extended behavioral capabilities, but it remains categorically identical to any other UE. There is no persistent infrastructure, no elevated vantage point, no fixed sensor installation, and no authoritative global knowledge of the coverage zone. The peer equality that makes NR-V2X Mode 2 architecturally elegant and deployment-flexible also makes it constitutively incapable of performing functions that require a categorically distinct entity.

Two failure modes need to be distinguished here, because resource expansion resolves only one of them. The first, the failure of base-station scheduling traced in Section~\ref{sec:3.2}, is capacity saturation --- insufficient aggregate bandwidth for the offered load --- which 3GPP's ongoing resource evolution directly addresses: flexible numerology (NR $\mu = 0$--3, slot durations 1 ms down to 0.125 ms), mini-slots, Massive MIMO, and carrier aggregation all raise the per-unit-time resource ceiling. The second is coordination failure, and it belongs to the peer-equal model: in specific generation intervals, safety-critical transmissions across all vehicles may simultaneously select conflicting channel resources, because a peer-equal allocation mechanism does not reliably break intra-class symmetry, regardless of how many channel resources are available. The simulation of Section~\ref{sec:7} gives this second failure mode a measured form. What decides whether a shared policy keeps safety vehicles apart is the size of the group it serves, not the number of subchannels: two safety vehicles on one shared policy separate at three, four and five subchannels alike, while five or more do not separate within the training budget (Section~\ref{sec:7.4}). The failure is therefore not caused by insufficient resources but by the inability of the allocation mechanism to guarantee distinct channel assignments to all safety-critical vehicles in every generation interval. 3GPP's resource evolution across Releases 15 through 19 raises the capacity ceiling. Mode 0 can establish a delivery floor.

The consequences of this peer equality become most visible in scenarios involving occluded or incomplete information --- scenarios that are not rare edge cases but routine features of urban traffic environments. Consider the ghost peeking scenario, a recognized class of occlusion-driven intersection hazard~\cite{ref21}: a pedestrian or vehicle is concealed behind a lateral obstruction --- a stationary vehicle, a delivery truck, a blind corner --- and emerges suddenly into the path of approaching traffic with insufficient warning time for evasive action. Mueller et al.~\cite{ref21} show that infrastructure-mounted sensors let automated vehicles handle such scenarios --- precisely the Seeing capability that defines the RCU's categorical advantage over road-level perception alone. Within the Mode 4/2 framework, the approaching vehicle receives no warning, because the hazard has no V2X presence until it has already emerged. Any vehicle equipped with forward-facing sensors may detect the obstruction and broadcast a warning BSM, but this warning faces two compounding limitations. First, the warning is issued from road level, where the sensor's line of sight to the occluded object is itself constrained by the same geometry that makes the hazard dangerous. Second, the broadcasting vehicle has no knowledge of which other vehicles within the zone need the warning and whether the warning has reached them with sufficient margin. Peer-to-peer sidelink communication cannot verify aggregate coverage across an entire traffic zone, because no individual UE possesses simultaneous knowledge of all other UEs' positions and reception states.

An RCU operating under the Mode 0 framework addresses this failure through three levels of information that are structurally inaccessible to peer UEs. At the first level, the RCU's elevated infrastructure sensors --- their specific composition a deployment choice rather than an architectural requirement --- detect objects occluded from road-level perspectives before they enter the traffic stream, enabling pre-emergence warning that no vehicle-mounted sensor can provide. At the second level, the RCU aggregates BSMs from all vehicles within its zone and correlates them with its own sensor data, building a complete kinematic picture of the coverage area that no individual UE can construct from local sensing alone. At the third level, when a UE-originated warning has been transmitted via sidelink, the RCU assesses whether the warning has reached all vehicles whose trajectories place them within the risk envelope of the identified hazard. Where coverage is insufficient, the RCU re-broadcasts the warning at infrastructure-grade power and priority, extending the effective safety envelope beyond what peer-to-peer transmission achieves.

It is important to be precise about the nature of this intervention. Mode 0 advisories, including hazard warnings, are non-mandatory in their default form. The RCU does not issue commands; it issues recommendations. Vehicles retain full autonomy over their own transmission decisions. This advisory default is entirely consistent with how the 3GPP mode framework operates more broadly. The conditions under which Mode 0 advisory authority may be escalated to mandatory status are specified in Section~\ref{sec:5.2} and motivated by the traffic risk taxonomy in Section~\ref{sec:3.4}. In the default case, vehicles have good reason to follow the RCU's advisories, because they are based on sensor data and global zone knowledge that vehicles cannot replicate independently.

\subsection{Institutional Authority and the Boundary of Advisory Sufficiency}\label{sec:3.4}

Beyond scheduling saturation and informational blindness lies a third structural limitation, one that emerges not from resource contention or sensing capability but from institutional mandate. There exists a specific class of traffic risk for which neither Mode 3/1 base station scheduling nor Mode 4/2 UE autonomy possesses the authority to act --- not because either lacks a technical means of communicating a warning, but because the scenario class in question requires an authority to compel compliance that neither entity holds. Identifying this class precisely defines the third failure mode this paper addresses.

We categorize all scenarios that challenge road traffic safety into three classes, differentiated by their scope of impact and the nature of the entities involved.

\textbf{Categories A and B --- Small-scope contact risk} encompass incidents whose risk envelope is confined to a single collision event that will not, if it occurs, cause lane blockage or trigger cascading secondary incidents. Category A involves two active traffic participants --- vehicle-to-vehicle collisions, vehicle-to-pedestrian collisions, and the ghost peeking scenario analyzed in Section~\ref{sec:3.3} are canonical examples. Category B involves one active participant and one passive physical entity in the traffic environment --- road debris, a fallen obstacle, an object intruding into a single lane without blocking it entirely, or a stalled vehicle that does not interrupt through-traffic. In both categories, the RCU's role is identical: its elevated observation layer detects the developing risk before any UE can, its communication layer issues warnings to all parties within the risk envelope, and its global zone awareness ensures coverage completeness. If prevention fails and an incident occurs, the RCU transitions to documentation and post-incident support: its sensor fusion record provides objective evidence for liability determination, and its communication layer coordinates emergency service notification. Throughout both categories, the advisory model is architecturally sufficient --- individual vehicle compliance with RCU warnings is rational, and failure to comply results in a localized incident whose scope does not expand independently of individual behavior.

\textbf{Category C --- Large-scope contact risk} is categorically different from both preceding classes in ways that have direct architectural implications. This class encompasses environmental hazards that, once they have produced an initial incident, create conditions for continuous cascading secondary incidents that cannot be halted by individual vehicle behavior and will not resolve until the physical environment changes. Road surface damage in a blind approach corridor, sudden dense fog in a tunnel exit zone, flash freezing on an elevated interchange, a major structural collapse blocking an entire carriageway --- these are the canonical Category C scenarios. Their defining characteristic is temporal escalation: each moment that passes without halting all approaching traffic toward the hazard zone produces a new secondary incident. The accident is not a discrete event but an ongoing process whose damage accumulates at a rate determined by traffic flow.

In Category C scenarios, neither the eNB/gNB nor any collection of vehicle UEs is architecturally capable of halting the escalation. The eNB/gNB lacks zone-specific hazard awareness, lacks the traffic authority mandate to issue stop commands, and cannot distinguish which of its connected UEs are approaching the hazard zone within the relevant time window. Individual vehicle UEs can detect the hazard only after entering proximity with it --- which may already be within the collision envelope --- and can broadcast warnings to nearby vehicles, but no individual UE possesses the global zone awareness to verify that all approaching vehicles within the warning horizon have received and acted upon the warning. Peer-to-peer warning propagation requires multiple relay hops and introduces latency that compounds with each hop, precisely when the time window is tightest.

Addressing Category C requires a specific combination of capabilities that no entity in the existing framework possesses simultaneously: elevated sensing that detects the hazard independently of V2X reports from affected vehicles, global zone awareness that identifies every approaching vehicle within the stop-distance horizon simultaneously, communication authority that reaches all vehicles in the zone through a single infrastructure broadcast, and institutional position within the traffic management domain whose mandate is precisely the protection of life and health of traffic participants. Each of the base station and the vehicle UE satisfies at most two of these four conditions; neither satisfies all four.

This gap creates what we term the \textbf{Category C boundary condition}: in large-scope contact risk scenarios that have already produced an initial incident, advisory authority is insufficient not because individual vehicles are unwilling to comply, but because the time window within which compliance must be achieved is shorter than the decision and response cycle of individually rational actors. A driver who receives an advisory warning about a hazard two seconds ahead must still perceive, decide, and execute --- and that process alone can exhaust the very time margin the advisory was meant to preserve. The advisory model presupposes that the compliance decision is available within the safety margin. In Category C scenarios, that presupposition fails, and neither existing entity class has any mechanism --- technical or institutional --- to compel the compliance that advisory recommendation cannot achieve in time. Section~\ref{sec:4.1} argues why the roadside is the correct location to resolve this gap, and Section~\ref{sec:5.2} specifies the resulting escalation mechanism.

The categorical gap that Sections~\ref{sec:3.2} through~\ref{sec:3.4} have traced from three independent directions --- scheduling, information, and authority --- is now precisely defined. What all three point toward is the same missing entity: fixed infrastructure rather than a mobile terminal, answerable to a traffic authority rather than a network operator, and bounded to a single traffic node rather than a general cellular coverage area or an isolated vehicle's sensing range. It possesses elevated sensing and persistent global zone knowledge that neither the base station nor any UE holds, and it operates under a safety mandate that neither shares. The existing framework contains exactly two entity classes. A third is necessary, and the evidence that industry is already deploying it --- without a standards home to inhabit --- is the subject of Section~\ref{sec:4.3}.

\section{The Case for Mode 0: Why a New Category Is Necessary}\label{sec:4}

\subsection{The Compute Sinking Argument: Why the Roadside Is the Correct Architectural Boundary}\label{sec:4.1}

The failure modes identified in Section~\ref{sec:3} share a common structural origin: the entities authorized to perform scheduling decisions are too far --- architecturally, institutionally, and in some cases physically --- from the safety-critical events they are meant to serve. The base station is distant from the traffic node in all three senses. The vehicle UE is physically proximate but informationally isolated, lacking the global zone awareness that effective safety scheduling requires. The question that follows is not whether to bring decision-making authority closer to the point of need --- the latency arithmetic of Section~\ref{sec:3.2} makes this unavoidable --- but precisely where the correct boundary lies.

Two candidate boundaries present themselves. The first is the operator's edge cloud, instantiated in 5G networks as Multi-Access Edge Computing (MEC) infrastructure deployed at or near the base station. The second is the traffic node itself --- the intersection, the merge point, the approach corridor --- where the safety-critical events physically occur. Both boundaries reduce latency relative to centralized cloud coordination. They are not equivalent, however, and the difference is decisive for safety-critical V2X scheduling.

The operator edge cloud reduces the geographic distance between the scheduling decision and the base station, but it does not eliminate the fundamental path: vehicle sensing event $\rightarrow$ uplink transmission $\rightarrow$ base station $\rightarrow$ edge cloud $\rightarrow$ scheduling decision $\rightarrow$ downlink grant $\rightarrow$ vehicle transmission. Each hop in this chain contributes latency variability. It is not average latency that determines whether a 20 ms pre-crash requirement is met --- it is worst-case latency, which is dominated by variability. A multi-hop path through operator infrastructure cannot provide the tight worst-case latency bounds that safety applications require, because the variability of each intermediate hop compounds across the chain.

The roadside boundary eliminates all intermediate hops for the M0 scheduling decision. The RCU is co-located with the sensors that detect the safety event, the vehicles that need to be warned, and the communication resources that carry the warning. The scheduling advisory is generated locally from local sensor data and broadcast locally over a bounded zone. The path from sensing event to advisory delivery is a single hop, and its latency is bounded by the RCU's local processing time, the time to gain access to a subchannel, and the air interface propagation delay. The first is under the control of the traffic authority that operates the RCU, and the last is short and predictable. Channel access is not deterministic, because the advisory shares the M0 pool with other safety traffic --- the contention that Section~\ref{sec:7} measures. This is nonetheless not merely faster than the edge cloud path. With no intermediate hops, variability no longer compounds across stages, so the worst case can be bounded by the design of the RCU and the size of the M0 pool rather than left to the state of several operator-controlled stages.

The institutional dimension of the boundary argument is equally important. The operator edge cloud is owned and optimized by the network operator, whose mandate is service quality across all subscribers and whose regulatory framework is telecommunications law. The roadside traffic node is owned and optimized by the traffic management authority, whose mandate is traffic safety and flow and whose regulatory framework is transportation law. Placing safety-critical scheduling authority within operator infrastructure means that safety optimization is always subordinate to, or at best competing with, the operator's primary objective of network performance across a general subscriber population. This is the institutional expression of the M0/M1 coupling problem identified in Section~\ref{sec:3.2}: not merely that safety traffic shares resources with entertainment and non-safety traffic, but that the entity controlling those resources has no primary mandate to prioritize safety. The RCU resolves this misalignment by placing scheduling advisory authority within traffic management infrastructure, where the sole optimization objective is safety and flow and M1 traffic is outside the mandate entirely. This is a hard architectural guarantee, not a soft organizational preference, that the safety objective will not be traded against competing performance objectives at the infrastructure level. Section~\ref{sec:4.3} shows that traffic authorities in several jurisdictions have reached the same boundary independently.

The authority gap identified in Section~\ref{sec:3.4} --- the Category C boundary condition --- has a specific architectural resolution: an emergency escalation protocol by which the RCU's advisory authority is elevated to mandatory status under precisely defined trigger conditions, activated by a verifiable Category C event, and subject to authorization by the traffic management authority (Section~\ref{sec:5.2}). The advisory default remains correct for Categories A and B and all non-emergency conditions. In Category C scenarios, the argument for Mode 0 is one of non-substitutability within the 3GPP V2X framework: of its entity classes, only the RCU simultaneously observes the zone, reaches all approaching vehicles, and operates under the traffic safety mandate. Traffic authorities already halt traffic by other means --- variable message signs, lane-control signals, tunnel barriers, and roadside-originated Decentralized Environmental Notification Messages (DENM) --- and Mode 0 does not replace them. It addresses the locations that lack such equipment and the hazards that develop faster than an operator can set a sign, and it lets the V2X message itself carry the authority to be treated as binding by the vehicle's application. This is the strongest form of the non-substitutability argument for Mode 0 standardization.

\subsection{The M0/M1 Information Classification and the Demand Separation Imperative}\label{sec:4.2}

The resource contention argument of Section~\ref{sec:3.2} established that M0 and M1 traffic compete for the same physical resources under the same scheduler. This framing, while correct, understates the structural severity of the problem. The deeper issue is not that competition exists --- it is that the two traffic classes have fundamentally asymmetric demand growth trajectories, which means that any architecture that couples their resource pools will produce systematically worsening M0 service quality over time, independent of how much supply is added.

We formalize this observation through a two-class information taxonomy, consistent with the service-level-requirement-based use case classification established in C-V2X industry guidance~\cite{ref22}.

\textbf{M0 --- Safety-critical traffic} comprises all information whose delayed or failed delivery has direct consequences for physical safety of traffic participants. This class includes Cooperative Awareness Messages and Basic Safety Messages transmitted at 1--10 Hz per vehicle as specified in ETSI EN 302 637-2~\cite{ref2}, pre-crash sensing warnings with end-to-end latency requirements of 20 ms or below as established in 3GPP TR 22.885~\cite{ref1}, intersection collision warnings, emergency vehicle preemption alerts, and RCU-originated zone advisories as defined in Section~\ref{sec:5}. M0 traffic is characterized by strict and non-negotiable latency bounds, relatively small per-message payload sizes (200--800 bytes per message per Car2Car C2CCC\_TR\_2052~\cite{ref23}), and a demand growth rate that is slow and bounded. A vehicle cannot meaningfully generate safety-relevant state updates faster than its own kinematic state changes, which is bounded by the physics of vehicular dynamics, and the 10 Hz CAM transmission ceiling specified in ETSI EN 302 637-2 reflects that bound for awareness messages. M0 demand is not fixed: safety-class services such as collective perception, vulnerable-road-user messages and maneuver coordination add traffic and may use larger messages. Their growth is tied to standards and deployments, however, whereas M1 demand grows continuously with consumer behavior. M0 demand per vehicle therefore grows slowly and in steps, in total approximately linearly with vehicle density, and it is far less sensitive to consumer trends than M1 demand.

\textbf{M1 --- Non-safety traffic} comprises all information whose delayed delivery degrades user experience but carries no direct safety consequence. This class includes in-vehicle infotainment, navigation and map updates, software-over-the-air downloads, passenger device video streaming, high-fidelity audio, and social media access. M1 traffic is characterized by high and rapidly growing data volume, tolerance for variable latency, and a demand trajectory driven simultaneously by increasing per-device consumption rates and increasing device density. The proliferation of large-format in-vehicle displays, the transition to higher-resolution streaming formats, and the growing expectation of always-on connectivity for vehicle occupants are not speculative future trends --- they are observable market dynamics already reshaping cellular network load profiles~\cite{ref24}.

\begin{figure}[!t]
\centering
\includegraphics[width=\figcolw,keepaspectratio]%
                {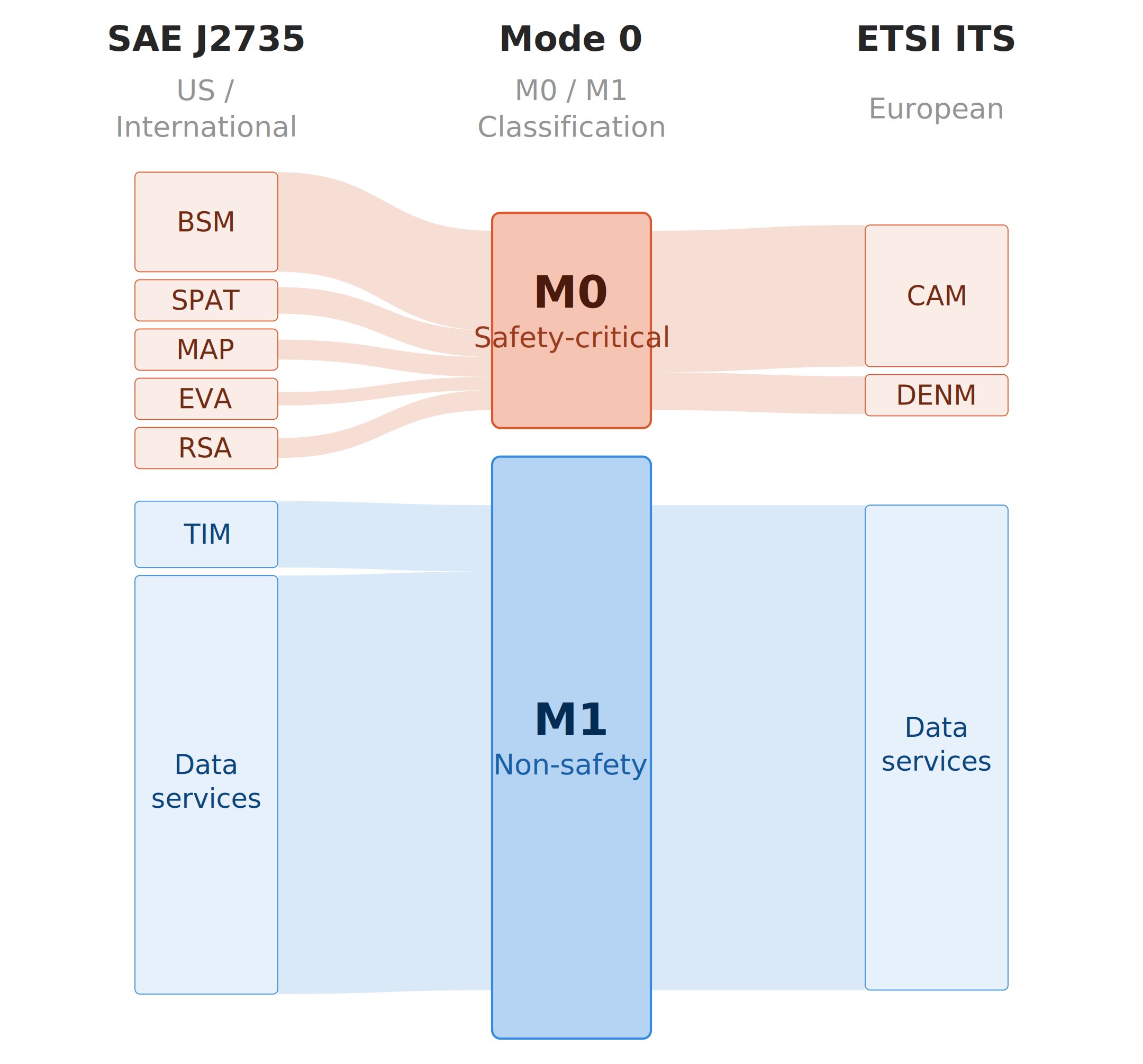}
\caption{M0/M1 traffic classification Sankey diagram across SAE J2735 (US/International, left) and ETSI ITS (European, right). M0-class: Basic Safety Message (BSM), Signal Phase and Timing (SPaT), MAP data (intersection geometry), Emergency Vehicle Alert (EVA), Road Side Alert (RSA) (SAE J2735~\cite{ref26}); Cooperative Awareness Message (CAM) (ETSI EN 302 637-2~\cite{ref2}) and DENM (ETSI EN 302 637-3~\cite{ref25}). M1-class: TIM, data services (both standards). RSA and DENM may carry M1-class content in non-urgent contexts. Flow heights encode per-vehicle throughput proportions derived from message specifications~\cite{ref2,ref23,ref25,ref26}; they do not represent vehicle population fractions. ``Data services'' volume exceeds all M0 messages by 2--3 orders of magnitude, per the analysis in Section~\ref{sec:4.2}.}
\label{fig:2}
\end{figure}

The volume asymmetry illustrated in Fig.~\ref{fig:2} reflects this analytical relationship --- flow heights are derived from message specifications (ETSI EN 302 637-2~\cite{ref2}, ETSI EN 302 637-3~\cite{ref25}, SAE J2735~\cite{ref26}, and Car2Car C2CCC\_TR\_2052~\cite{ref23}) rather than empirical measurement. The Sankey encodes data throughput proportions per vehicle, not vehicle population fractions: a single vehicle's M0 stream (CAMs at 10 Hz, 200--800 byte payloads) consumes approximately 16--64 kbit/s, about 30 kbit/s at the average CAM size of roughly 350 bytes reported in~\cite{ref23}, while a vehicle carrying two passengers streaming standard-definition video generates 2--20 Mbit/s of M1 traffic --- roughly two to three orders of magnitude larger at typical message sizes. The multidimensional possibilities of multiple passengers per vehicle and multiple terminals per passenger could further widen this order-of-magnitude gap --- an effect already observable in the V2P context discussed in Section~\ref{sec:8.5}. Critically, M0 and M1 demand are not independent: population density drives both vehicle counts (M0 demand) and cellular subscriber density (M1 demand, compounded by in-vehicle occupant devices), meaning peak M0 safety-critical periods --- rush hour at high-density urban intersections --- coincide with peak M1 throughput demand at exactly the locations and moments where coordination failure is most dangerous. This demand correlation makes the shared-pool architecture doubly problematic: it fails hardest precisely when it matters most.

\textbf{A note on orthogonality.} The M0/M1 classification is defined by the safety-criticality and latency requirements of the information --- not by communication link type (vehicle-to-vehicle, V2V; vehicle-to-infrastructure, V2I; vehicle-to-network, V2N) or transport interface (PC5, Uu). These three dimensions are orthogonal: a BSM transmitted via a cloud-relay Uu path retains its M0-class safety requirements; a Traveler Information Message (TIM) delivered over PC5 from an RSU may carry M1-class information by latency tolerance. Neither SAE J2735 nor ETSI EN 302 637-2/3 normatively restricts specific message types to specific interfaces, and Mode 0 inherits this interface-agnosticism deliberately. The M0 resource pool specification operates at the resource management layer; which transport mechanism carries the traffic is a deployment-specific decision that Mode 0 does not constrain. This deliberate non-binding preserves compatibility with hybrid deployment architectures --- Uu-relay C-V2X, PC5-primary Mode 4/2, and future air-interface configurations --- without requiring Mode 0's resource management principles to be revised as transport mechanisms evolve.

The simulation of Section~\ref{sec:6.3} uses a balanced vehicle population composition of 50\% M0-class and 50\% M1-class (\texttt{m0\_ratio} = 0.50), chosen to test symmetric fleet conditions; this vehicle count ratio is distinct from the data throughput asymmetry illustrated in Fig.~\ref{fig:2}. For simplicity, the simulation also treats each vehicle as generating exclusively M0 or exclusively M1 traffic, whereas --- consistent with the orthogonality established above --- a real vehicle generates both simultaneously; Section~\ref{sec:6.3} details this simplification and its scope. 

The asymmetry between M0 and M1 demand growth trajectories is the central structural problem that Mode 0 addresses. Supply-side solutions --- additional spectrum, more granular resource block allocation, higher-order modulation schemes --- address demand/supply ratio degradation by increasing supply. They cannot, however, provide a permanent architectural solution to demand asymmetry, because each supply increase is a discrete event while M1 demand growth is continuous. Every generation of cellular technology has confronted this pattern: the supply increase of a new air interface standard is eventually consumed by demand growth, motivating the next generation. For general cellular services, this cycle is an acceptable engineering and economic reality. For M0 safety-critical traffic, it is not acceptable, because each degradation episode within the cycle carries real safety consequences.

The correct solution operates on the demand side rather than the supply side, and specifically on the coupling between demand classes. If M0 and M1 traffic are separated at the infrastructure level --- each class assigned to a dedicated resource pool managed by the infrastructure entity whose mandate aligns with that class's requirements --- then M1 demand growth no longer competes with M0 service quality. The M0 resource pool, managed by the RCU, serves a demand that grows slowly and predictably. The M1 resource pool, managed by the base station, serves a demand that grows rapidly but is delay-tolerant --- precisely the condition that base station schedulers are optimized to handle through aggregate throughput maximization. The practical mechanism by which the two infrastructure domains coordinate at this boundary --- beyond the resource-pool separation itself --- is an open question, taken up in Sections~\ref{sec:8.2} and~\ref{sec:8.9}.

It is important to address a potential objection directly: does routing M1 traffic to the base station merely transfer the congestion problem, degrading M1 service quality in exchange for protecting M0? The base station was designed for high-volume, delay-tolerant, aggregate-optimized traffic across a large subscriber pool --- which is precisely the description of M1 traffic. Removing M0 from the base station's resource pool gives the base station scheduler a homogeneous M1 workload that it can optimize without the constraint of serving heterogeneous traffic classes with conflicting latency requirements. Mode 0 routes each traffic class to the infrastructure entity optimized to serve it. Whether a separate M0 pool helps depends on the load, and the deployment rule that the simulation supports is therefore conditional (Section~\ref{sec:7.6}): the RCU reserves a pool equal to the safety demand where spectrum is spare --- a pool that can be reused across kinematically non-conflicting sub-zones, as in the spatial reuse mechanism of Section~\ref{sec:5.2} --- and keeps the pool shared at the critical load. This is a principled safety-class prioritization rather than a Pareto improvement, and it is the correct trade for a safety architecture whose mandate is life and health protection.

\subsection{Existing Deployments as Evidence for the Gap}\label{sec:4.3}

A theoretical argument for a new standards category is necessarily prospective: it claims that an entity which the current framework does not accommodate is needed and should be standardized. The argument of Sections~\ref{sec:3.2} through~\ref{sec:3.4} makes this case on structural grounds. A stronger form of the same argument is available when independent real-world deployments have already converged on the proposed entity without a standards framework to guide them. Convergent independent discovery is strong evidence that the entity responds to a genuine and universal architectural need.

This evidence is available for the RCU. Multiple independent deployment ecosystems, proceeding under different regulatory frameworks and driven by different institutional actors, have arrived at an entity that matches the Mode 0 RCU specification in its essential characteristics: fixed roadside deployment, computing capability for local traffic information processing and decision-making, ownership by traffic management authorities rather than network operators, and a service scope bounded by traffic participants rather than general cellular subscribers.

The most fully developed regulatory expression of this convergence is found in China's vehicle-road-cloud integration standards ecosystem. DB11/T 2329.1-2024~\cite{ref7}, a Beijing Municipal Standard effective from March 2025, defines the Roadside Computing Unit (RCU) --- literally ``Roadside Computing Unit'' in Chinese, though the standard maps this to the English acronym MEC --- as ``a computing module, device, or facility deployed along roads, highways, or at venues that, in conjunction with other facilities or systems, performs traffic information aggregation, processing, and decision-making.'' We note that the standard's English rendering as ``MEC'' is terminologically imprecise for our purposes: MEC in the telecommunications literature refers to a broad infrastructure category that includes edge cloud deployments at or near base stations, whereas the entity in DB11/T 2329.1-2024 is specifically bounded to roadside deployment with traffic management authority ownership. The literal translation captures this specificity more accurately than the standard's own English rendering, which is why we adopt RCU as the precise term throughout this paper. The standard distinguishes explicitly between the RSU --- a communication relay device --- and the MEC roadside computing unit, recognizing them as separate entities within the roadside infrastructure layer. This standard is operational, not aspirational: it specifies data exchange protocols, device identification naming conventions, and communication requirements for a device actively being procured and deployed at scale within Beijing's Advanced Autonomous Driving Demonstration Zone.

T/ITS 0224.1-2025~\cite{ref8} is an industry-alliance group standard, issued by the China Intelligent Transportation Industry Alliance and effective from July 2025; it provides a functional architecture specification for the same entity class. The standard explicitly positions the edge computing facility within the edge subsystem of the vehicle-road cooperative system at roadside deployment, and defines a layered operating system architecture comprising hardware abstraction, computing and communication middleware, service framework, and application framework. Fusion computation is confirmed as a core service function, and data services include real-time fusion computation result output with transmission to RSUs or IoV platforms.

A parallel and institutionally distinct development is occurring simultaneously on the network operator side. China Unicom's Smart Network Innovation Center announced at the 2023 China 5G+ Industrial Internet Conference the release of the Roadside Multi-Access Edge Computing Unit (RS-MEC)~\cite{ref9}, jointly developed with Lenovo Group. The RS-MEC carries high-performance GPU hardware, supports 5G/V2X dual-mode communication, and provides structured processing of roadside perception data, precise identification of traffic participants, and real-time intelligent decision-making for the operator's service operations. Critically, the RS-MEC is incorporated into China Unicom's unified MEC management architecture --- it is operator-owned infrastructure deployed at the roadside boundary, optimized for the operator's service delivery objectives. This makes the RS-MEC the operator-side parallel to the traffic authority's RCU: two roadside computing entities converging on the same physical location from opposite institutional domains.

The co-existence of these two deployments at the same traffic nodes in China today reveals the urgency of Mode 0 standardization with a precision that theory alone cannot provide. The traffic authority's RCU and the operator's RS-MEC may literally be co-located at the same Beijing intersection, with no standardized protocol governing how they coordinate on M0 resource pool allocation. The traffic authority needs to know which communication resources are reserved for M0 safety communications; the operator needs to know which communication resources are available for its RS-MEC's 5G/V2X services. Without adequate policy support for Mode 0 (such as the cross-domain configuration interface outlined in Section~\ref{sec:8.2}), this coordination is achieved, if at all, through ad-hoc bilateral agreements that cannot scale across the thousands of intersections that large-scale deployment requires. Mode 0 standardization is not a proposal that roadside computing infrastructure should exist. It already exists, from both sides of the institutional boundary. Mode 0 is the interface specification that makes it interoperable.

The significance of these citations must be stated precisely. We are not arguing that 3GPP should adopt Chinese standards. We are arguing that the industry --- both traffic authorities and network operators --- has independently arrived at entities matching the RCU and RS-MEC specifications, driven by the same underlying architectural need, and that the absence of a 3GPP mode definition creates interoperability risk that grows with each additional deployment.

Parallel developments in other major vehicle markets reinforce the universality of this convergence. European C-ITS infrastructure programs, documented by Kotsi et al.~\cite{ref27} and Sjoberg et al.~\cite{ref28}, have progressively expanded roadside ITS station functional scope toward local processing and cooperative perception aggregation. United States C-V2X deployment programs, advancing under USDOT guidance~\cite{ref29}, have similarly emphasized roadside unit intelligence as a key enabler of safety application performance at complex intersections.

\section{Formal Definition of Mode 0 and the RCU Entity Class}\label{sec:5}

\subsection{Entity Classification: What the RCU Is and Is Not}\label{sec:5.1}

The arguments of Sections~\ref{sec:3} and~\ref{sec:4} converge on a single structural conclusion: the entity needed to close the gap in the 3GPP V2X mode framework is neither a base station nor a UE, and cannot be accurately represented as a variant of either. Before specifying what Mode 0 is, it is necessary to establish this negative case with precision, because the temptation to classify the RCU as an extended RSU, a specialized MEC node, or a Mode 2(d) group leader with additional capabilities will be the natural reflex of anyone working within the existing taxonomy. Each of these classifications is wrong, and the wrongness is not terminological --- it is architectural.

\textbf{The RCU is not a base station.} The eNB and gNB are general-purpose cellular infrastructure entities whose defining characteristics are: service to a subscriber population defined by network operator contract rather than by participation in a specific physical activity; optimization objectives oriented toward aggregate network performance metrics; ownership and operation by network operators under telecommunications regulatory frameworks; and deployment geometry determined by cellular coverage planning. The RCU shares none of these characteristics. Its service objects are exclusively entities participating in traffic behaviors within a geographically bounded zone. Its optimization objective is traffic safety and flow, not network performance. It is owned and operated by a traffic management authority under transportation regulatory frameworks. Its deployment geometry is determined by traffic node geography, not cellular coverage planning.

\textbf{The RCU is not a UE.} The UE in 3GPP's framework is defined by mobility, by subscription to a network operator's service, and by the peer equality that characterizes all UE-class entities regardless of behavioral sophistication. As established in Section~\ref{sec:3.3}, the Mode 2(d) group leader is explicitly not a different class or category of UE~\cite{ref20}. The RCU violates each of these defining characteristics. It is fixed infrastructure, not a mobile terminal. It does not subscribe to a network operator's service in the manner of a UE. And it is categorically not a peer of the vehicles it serves: it possesses elevated sensor installations, global zone awareness, and advisory authority that no vehicle UE can replicate.

\textbf{The RCU is not adequately described as an RSU with added compute.} The traditional RSU is a communication relay device whose primary function is message forwarding. Adding computing capability to an RSU produces a device that can process messages locally before forwarding them --- a meaningful capability extension, but not a new entity class. The RCU's role is categorically different: it exercises advisory scheduling authority over a bounded traffic zone, performs multi-modal sensor fusion to detect hazards that V2X communication cannot reveal, validates warning coverage adequacy across its zone, and serves as the institutional boundary between traffic management authority and network operator domains. We adopt the term RCU rather than RSU-MEC to mark this categorical distinction, noting that the Chinese standard DB11/T 2329.1-2024~\cite{ref7}, while mapping Roadside Computing Unit (RCU) to the English acronym MEC in its own translation, uses the Chinese term that translates literally as Roadside Computing Unit --- a translation that captures the entity's nature more accurately than the standard's own English rendering.

\textbf{The Seeing, Speaking, and Thinking framework.} The RCU's categorical distinction from all existing entity classes can be articulated along three functional dimensions. It must be capable of \textbf{Seeing} --- observing the traffic zone from a vantage point inaccessible to road-level sensors. It must be capable of \textbf{Speaking} --- communicating with vehicle UEs over PC5 sidelink and with management platforms via backhaul. It must be capable of \textbf{Thinking} --- fusing sensor and V2X data into a global zone model that supports advisory computation. The eNB/gNB possesses Thinking and Speaking without Seeing. The UE possesses all three but at bounded Thinking capacity and road-level Seeing. The RCU alone possesses elevated Seeing that neither can replicate.

\textbf{The RCU as infrastructure ensemble.} Having established what the RCU is not and characterized it functionally, we can now state what it is with the structural specificity that a standards proposal requires. The RCU is most precisely understood as an infrastructure ensemble rather than a single device: three functionally distinct components, one for each dimension above, are necessary, and together with traffic-authority ownership (below) sufficient, for an entity to exercise the RCU role.

The first component is the \textbf{observation layer} (Seeing): elevated sensing hardware deployed at the traffic node, which may include cameras, millimeter-wave radar, LiDAR, and other exploitable sensor types, positioned at intersection height or on elevated infrastructure. The observation layer is responsible for detecting all traffic participants within the zone --- vehicles, pedestrians, cyclists --- including objects occluded from road level by parked vehicles, buses, or other obstructions. This is the component that makes possible both a comprehensive regional mobility model and pre-emergence hazard detection --- the RCU's most fundamental categorical advantages over vehicle UEs: it observes what no mobile terminal can observe, because its elevation and fixed positioning give it a geometric vantage point that mobility permanently denies to any vehicle-mounted sensor. The same continuous, zone-wide vantage point that catches an occluded hazard the moment it emerges also accumulates, over time, into the comprehensive picture of vehicle flows and densities that a regional mobility model requires --- a byproduct no single vehicle's transient, self-centered view could ever assemble.

The second component is the \textbf{communication layer} (Speaking): a roadside unit at the traffic node responsible for communication on two fronts. Toward vehicles, it exchanges V2X messages over the PC5 sidelink interface --- broadcasting Mode 0 advisory messages (sub-zone resource recommendations and hazard re-broadcasts) and aggregating BSM and CAM inputs from all vehicles within the zone. Toward the network operator's infrastructure and any broader traffic management platform, it exchanges status, configuration, and coordination data via backhaul. This component corresponds to what the RSU performs in existing V2X infrastructure, extended to cover both directions. Its presence is necessary but not sufficient for the RCU role.

The third component is the \textbf{evaluation layer} (Thinking): a local MEC facility at the traffic node that serves as the ensemble's computational brain. The evaluation layer fuses the observation layer's sensor data with the communication layer's aggregated V2X inputs to construct and continuously update the global zone model --- the complete kinematic picture of all traffic participants --- that is the information substrate for advisory scheduling. It executes the zone partitioning logic, computes sub-zone resource recommendations, assesses warning coverage adequacy, determines when and whether to re-broadcast UE-originated warnings, and --- in Category C scenarios as defined in Section~\ref{sec:3.4} --- assesses whether the conditions for emergency escalation have been met.

The RCU is the integrated operation of all three layers as a unified system under traffic management authority ownership and governance. This is why the RSU-MEC compound is an insufficient characterization: it describes two of the three components and omits the observation layer, the defining differentiator between the RCU and all predecessor concepts, which lets the ensemble perceive what vehicles cannot perceive.

\textbf{A fourth distinguishing dimension --- institutional authority.} The three functional layers define what the RCU \emph{can} do; a fourth dimension defines what it is \emph{mandated} to do. The RCU is owned by a traffic management authority whose legal and regulatory mandate is traffic safety and flow --- not network throughput. This institutional position is the fundamental distinction between the RCU and all prior roadside infrastructure concepts: an RSU or an RSU-MEC compound may acquire equivalent technical capabilities, but without the traffic management authority's mandate they cannot exercise mandatory escalation authority. It is this mandate --- not the hardware --- that transforms advisory capability into enforceable safety authority when Category C conditions are met. The three functional layers are the \emph{means}; the institutional authority is the \emph{standing} that makes the means consequential.

\textbf{Performance variability across deployments.} The Mode 0 specification defines what the RCU role must be capable of doing; it does not prescribe the hardware, algorithmic parameters or coordination arrangements through which a deployment achieves it. Deployment quality varies along three dimensions, one for each part of the ensemble. Hardware integration quality is the sensing resolution and coverage of the observation layer, the reliability and range of the communication layer, and the throughput and latency of the evaluation layer. Algorithmic effectiveness is the design of the zone partitioning logic, the mobility assessment, the scheduling advisory algorithm, the warning coverage assessment and the M0/M1 resource pool sizing, which different traffic authorities will tune to local traffic patterns, intersection geometry and safety priorities. Cross-domain coordination quality is the effectiveness of the configuration interface between the traffic management authority and the network operator on M0 resource pool allocation, as discussed in Section~\ref{sec:8.2}. The specification defines the professional position and its essential requirements; how well a given deployment inhabits that position is the work of traffic management authorities.

One boundary condition in the RCU entity classification deserves explicit acknowledgment. The current Mode 0 specification defines the RCU's service objects as entities participating in traffic behaviors within its zone, a definition that encompasses vehicle UEs straightforwardly. Pedestrians and cyclists on the road are equally within this scope. However, a pedestrian's personal device is simultaneously a service object of the RCU in the traffic safety domain and a subscriber device of the eNB/gNB in the telecommunications domain, creating a dual-domain membership that the binary entity taxonomy does not resolve. The vehicle UE has no equivalent dual membership: its communication behavior in the traffic zone is entirely mediated through its on-board unit (OBU) within the Mode 0 framework. The pedestrian's device, by contrast, must be reachable through both domains for complete V2P safety coverage. Table~\ref{tab:1} reflects the primary entity class distinctions for the V2V-focused specification of Mode 0. The V2P extension and the dual-domain pedestrian problem are treated as open issues in Section~\ref{sec:8.5}.

\begin{table*}[!t]
\caption{Entity Class Comparison}
\label{tab:1}
\centering
\footnotesize
\renewcommand{\arraystretch}{1.20}
\setlength{\tabcolsep}{4pt}
\begin{tabular}{@{}>{\raggedright\arraybackslash}p{2.35cm}>{\raggedright\arraybackslash}p{2.75cm}>{\raggedright\arraybackslash}p{2.75cm}>{\raggedright\arraybackslash}p{6.0cm}@{}}
\toprule
\textbf{Dimension} & \textbf{eNB / gNB} & \textbf{UE} & \textbf{RCU (Mode 0)} \\
\midrule
Physical composition & Base station hardware & Mobile terminal & Ensemble: observation layer (Seeing) + communication layer (Speaking) + evaluation layer (Thinking) \\
Service objects & General cellular subscribers & Self / peer UEs & Traffic participants in bounded zone \\
Ownership & Network operator & Individual / fleet & Traffic management authority \\
Optimization objective & Network throughput and fairness & Individual utility & Traffic safety and flow \\
Mobility & Fixed (macro / small cell) & Mobile & Fixed (traffic node) \\
Sensor capability & Antenna arrays (no environmental sensing) & Road-level vehicle-mounted sensors & Elevated multi-modal (deployment-specific) \\
Zone awareness & Cell-wide aggregate & Local sensing only & Global within traffic zone \\
Scheduling authority & Mandatory grant-based & Autonomous / peer & Advisory default; mandatory escalation in Category C scenarios (Section~\ref{sec:5.2}) \\
Regulatory framework & Telecommunications law & --- & Transportation law \\
V2P service object scope & All Uu subscribers including pedestrian devices & Not applicable & Traffic zone participants; pedestrian device reachability is an open extension (Section~\ref{sec:8.5}) \\
\bottomrule
\end{tabular}
\end{table*}

\subsection{Mode 0 Operational Specification}\label{sec:5.2}

With the RCU entity class established, we define Mode 0 as the resource allocation mode in which an RCU provides zone-based advisory scheduling for M0 safety-critical V2X communications within its coverage zone. Before specifying the nine operational components that all compliant deployments must provide, numbered (1) to (9) below, we define the Mode 0 subfamily taxonomy that characterizes the capability contribution of vehicle UEs within the framework.

\textbf{Mode 0 Subfamily Classification.} The Mode 0 specification defines the RCU's role and the resource architecture that all compliant deployments must provide. The coordination quality that vehicle UEs contribute within this framework depends on whether those UEs possess sufficient on-board compute to operate as independent MARL agents --- a hardware capability that varies across vehicle generations and fleet compositions. The capability tier denotes a UE's ability to execute an individualized, locally-conditioned coordination policy; the per-vehicle MARL actor used in this paper's simulation is the reference implementation of that capability, not a normative requirement. Mode 0 therefore defines a family of subfamilies, analogous to the sub-modes 2(a) to 2(d) that the Release 16 study of NR-V2X Mode 2 considered~\cite{ref18}, specifying the UE capability tier for a given deployment context.

\textbf{Mode 0a --- All-Passive UEs.} No vehicle in the coverage zone has the hardware capability to operate as an independent per-vehicle MARL agent. Vehicle UEs receive the RCU's advisory recommendations and act on them through their existing resource selection logic, but they do not contribute individualized coordination policies. Mode 0a is the mandatory minimum compliance level for all Mode 0 deployments: it must function correctly even for fully legacy fleets in which the least-capable vehicle determines the coordination floor. This is the bucket-effect baseline --- the system's performance lower bound is set by the stave of minimum UE capability --- and Mode 0a is specified as the safety-class minimum regardless of that capability. The simulation of Section~\ref{sec:7} shows how that minimum can and cannot be met. A shared learned policy cannot carry it alone when it serves many vehicles: at the critical safety load it stays at the random-choice floor and delivers less than standardized sensing. The minimum therefore rests on the execution-side functions of the RCU --- grouping passive vehicles finely, so that each shared policy serves few of them, and a safety partition where spectrum allows (Sections~\ref{sec:7.4},~\ref{sec:7.6} and ~\ref{sec:7.7}). In the current 2026 traffic environment, where legacy fuel vehicles without redundant onboard compute coexist with connected vehicles, Mode 0a is the correct safety net: a standard that works for the fleet as it actually exists, not as it ideally would.

\textbf{Mode 0b --- Hybrid UEs.} A mixed fleet in which some vehicle UEs have the hardware capability to operate as per-vehicle MARL agents and others do not. Mode 0b is the expected transitional state during fleet modernization, as older passive vehicles coexist with newer active vehicles. The RCU identifies UE capability through the Mode 0 advisory exchange and adapts its scheduling recommendations accordingly, applying Mode 0a-compatible advisory logic for passive UEs and Mode 0c-compatible coordination incentives for active UEs.

\textbf{Mode 0c --- All-Active UEs.} All vehicles in the coverage zone have the hardware capability to operate as independent per-vehicle MARL agents. Mode 0c is the optimal long-term target: each vehicle runs its own policy network conditioned on its local observation, escaping the symmetric-Nash coordination floor --- the analytical collision floor under independent uniform selection (Section~\ref{sec:6.5.2}) --- which Mode 0a's shared-actor architecture stays at or near whenever one shared policy serves five or more safety vehicles (Section~\ref{sec:7.4}). Mode 0c coordinates subchannel choice at every load tested, and the simulation attributes its advantage over Mode 0a and over standardized sensing-based scheduling to this coordinated choice, not to power control (Section~\ref{sec:7.3}).

The simulation program evaluates all three tiers: Mode 0a's shared per-class policies separate small class groups but fall to the random floor at the critical safety load, below the standardized baseline; Mode 0c coordinates at every load tested; and Mode 0b keeps full safety delivery for passive and active vehicles alike (Sections~\ref{sec:7.2},~\ref{sec:7.4} and ~\ref{sec:7.7}). The operational specification below applies to all subfamily configurations; the subfamily determines how effectively vehicle UEs exploit the coordination opportunity the specification provides.

\textbf{(1) Zone definition.} The RCU defines a coverage zone corresponding to a traffic node and its approach corridors. Zone boundaries are determined by traffic geometry rather than radio propagation characteristics. Within the zone, the RCU maintains a persistent real-time model of all traffic participants' positions, velocities, headings, and trajectory predictions, derived from sensor fusion of its own sensor data and aggregated BSM/CAM messages received from vehicles.

\textbf{(2) Resource partitioning.} A subset of available PC5 sidelink subchannels is designated as the M0 resource pool, reserved for safety-critical communications within the zone. The remaining subchannels constitute the M1 resource pool, available for base station or operator infrastructure scheduling of non-safety traffic. The partition boundary is a deployment configuration parameter set by the traffic management authority based on local traffic density, safety application requirements, and available spectrum; Section~\ref{sec:7.6} tests the partition and Section~\ref{sec:8.9} gives the sizing procedure.

\textbf{(3) Advisory scheduling.} Within the M0 resource pool, the RCU divides the traffic zone into spatial sub-zones --- the minimum sub-zone being a single vehicle UE when kinematic precision warrants it --- based on kinematic parameters. The purpose of this division is to minimize M0 subchannel conflicts among vehicles that are kinematically likely to interact: vehicles on converging trajectories or sharing a road segment are assigned distinct subchannel recommendations to eliminate co-channel interference at moments of closest proximity. Vehicles are assigned to sub-zones according to their driving direction, speed, and trajectory, with the goal of ensuring that vehicles in adjacent sub-zones are assigned non-interfering subchannels within the M0 pool. The granularity of this division is itself a function of the number of UEs within the zone and the RCU's available processing and signaling budget: when UE count is low, granularity can straightforwardly approach one sub-zone per vehicle; as UE count grows, vehicles may have to be aggregated into coarser sub-zones to remain within that budget. The budget is not the cost of executing a per-vehicle policy, which Section~\ref{sec:7.9} measures in microseconds on the vehicle. It is the RCU's sensor fusion and conflict-graph computation, and the signaling needed to address every vehicle in every interval --- a cost the simulation does not charge (Section~\ref{sec:7.9}). This aggregation is not capability-blind: Mode 0c vehicles, whose per-vehicle actors can resolve within-sub-zone contention independently, tolerate coarser grouping, whereas Mode 0a vehicles --- which have no such compensating capability and depend entirely on the RCU's own separation --- benefit from finer, more conservative sub-zone assignment. The simulation gives this a measured basis: a shared policy separates a group of two safety vehicles in every seed, a group of three usually and slowly, and a group of five or more never, and in the mixed fleet the passive safety pair separated in every seed (Sections~\ref{sec:7.4} and~\ref{sec:7.7}). Passive vehicles should therefore be grouped finely, so that each shared policy serves few of them, while active vehicles need not be. The RCU's aggregation logic therefore weighs kinematic proximity, UE capability tier, and its own computational load jointly, rather than treating sub-zone granularity as a purely kinematic decision. The RCU broadcasts sub-zone resource recommendations to vehicles within its zone as advisory messages over the PC5 interface. This advisory mechanism differs fundamentally from the grant-based scheduling of Mode 3/1: the base station issues mandatory grants, while the RCU issues recommendations that vehicles may follow or disregard under normal operating conditions.

\textbf{(4) Spatial subchannel reuse.} Crucially, sub-zones that are kinematically unlikely to intersect --- separated by physical barriers, on diverging trajectories, or beyond plausible encounter range within the current planning horizon --- may be assigned the same M0 subchannel resources without conflict risk. This spatial subchannel reuse within the zone is the mechanism by which Mode 0 serves more vehicles than the raw M0 subchannel count $M_{m_0}$ might suggest: the effective reuse factor depends on zone geometry and traffic flow topology, both continuously available to the RCU's evaluation layer and inaccessible to any individual vehicle UE. The spatial reuse allocation is only constructible from the global kinematic picture --- this is a capability unique to the RCU entity class.

\textbf{Graph coloring interpretation.} Formally, the sub-zone assignment problem is a graph coloring problem: sub-zones are nodes, kinematic conflict relationships are edges, and subchannels are colors. Two limits of the formulation should be stated. First, a coloring gives one subchannel to each sub-zone, so it separates sub-zones from one another and not the vehicles inside a sub-zone: a sub-zone holding several vehicles is a single node, and those vehicles still need a rule of their own --- the finest sub-zone, a single vehicle, or the per-vehicle actors of Mode 0c. Second, the Four Color Theorem applies only to planar conflict graphs. When conflicts arise only between sub-zones that border one another on a single roadway plane, the graph is planar and $M_{m_0} \ge 4$ subchannels suffice, with sub-zones colored to maximize reuse. Interference, however, reaches beyond adjacent sub-zones, and each long-range conflict moves the graph away from planarity. The simulated pool is the extreme case: every vehicle can interfere with every other, so the five safety vehicles form a complete graph on five nodes, which is not planar and needs five colors --- a collision-free assignment of the safety vehicles needs five subchannels, as in Section~\ref{sec:7}. The planar bound is therefore a best case for zones whose conflicts are local, not a design rule. Conflict graphs with non-planar structure --- multi-level interchanges, long-range interference couplings --- are identified, together with formal reuse-factor analysis under dynamic mobility, as a direction for companion work.

\textbf{(5) Density-predictability alignment.} A favorable self-reinforcing property of the spatial reuse mechanism deserves note. Traffic flow theory distinguishes free flow (low density) from congested flow (high density). In free flow, vehicles exercise large behavioral degrees of freedom --- lane changes, overtaking, variable speeds --- making individual trajectory prediction uncertain over any meaningful planning horizon. In congested flow, vehicle trajectories become coupled: speeds equalize, lateral movements become rare, and the kinematic state of each vehicle is largely determined by its neighbors. This creates a favorable alignment for the reuse algorithm: when density is low, reuse demand is low but trajectory prediction is less reliable; when density is high, reuse demand is high but trajectory prediction is more reliable because congested flow constrains individual behavior. The conditions that make spatial reuse most valuable are also the conditions under which the algorithm is most accurate. The RCU's evaluation layer can adapt the planning horizon to instantaneous density --- tightening conflict boundaries at high density, widening them at low density --- as a straightforward operational parameter with no architectural implication.

\textbf{(6) Compliance model.} Mode 0 advisory compliance is non-mandatory in its default form. This is a deliberate architectural choice consistent with how the 3GPP mode framework operates across all existing modes. Non-mandatory compliance does not undermine Mode 0's effectiveness in Categories A and B scenarios, because compliance is, in the typical case, in each vehicle's own interest: the RCU's sub-zone assignments are derived from global zone awareness that no individual vehicle can construct from local sensing alone, so following the RCU's recommendation reduces the probability that the vehicle's own safety messages experience interference. Compliance is therefore plausibly an individually rational strategy, and it does not require protocol enforcement; the benefit grows with the share of vehicles that follow the recommendation. This is an argument, not a result: the simulation assumes compliance and does not test partial compliance (Section~\ref{sec:7.9}). The informational compliance model established here operates independently of any economic incentive structure. A complementary layer of economic compliance incentives is outlined in Section~\ref{sec:8.3}.

\textbf{(7) Advisory latency structure.} Unlike Mode 3/1's multi-hop scheduling cycle --- SR and buffer status report (BSR) uplink request, scheduling grant, downlink, then UE transmission --- or Mode 4/2's extended sensing-window pre-selection (a sensing window of 1000 ms in LTE Mode 4~\cite{ref17} and of 100 or 1100 ms, as configured, in NR Mode 2~\cite{ref10}), the Mode 0 advisory is a single PC5 broadcast hop from the RCU's evaluation layer to vehicle UEs, with no uplink grant overhead and no sensing window latency. This single-hop structure is the structural precondition for bounding worst-case advisory latency: the advisory delivery latency is bounded by the RCU's processing cycle and one PC5 propagation delay, not by any multi-party protocol negotiation. The RCU's fixed installation makes dedicated Global Navigation Satellite System (GNSS) hardware a natural, though not architecturally mandatory, choice for a precision timing reference for the zone --- a shared reference point that no individual vehicle's own receiver could serve as, given a vehicle's motion and its purely local vantage --- enabling synchronization at transmission-time-interval (TTI) resolution, which stabilizes the sidelink timing hierarchy, particularly in handover boundary regions where UE-negotiated synchronization is most uncertain.

\textbf{(8) Warning validation and re-broadcast.} When a UE-originated safety warning has been transmitted via sidelink, the RCU performs coverage adequacy assessment. Where the assessment identifies vehicles within the risk envelope that are unlikely to have received the warning, the RCU re-broadcasts the warning at infrastructure-grade power within the M0 resource pool. This re-broadcast is labeled as an RCU relay rather than a primary origination. As with all default Mode 0 advisories, the re-broadcast is non-mandatory.

\textbf{(9) Emergency escalation protocol.} Under normal operating conditions, all Mode 0 advisories are non-mandatory, and the advisory default is appropriate for Category A and B traffic risk scenarios as defined in Section~\ref{sec:3.4}. Category C scenarios --- large-scope environmental hazards producing cascading secondary incident risk --- introduce a boundary condition under which the advisory model is architecturally insufficient and an escalation mechanism is required. Mode 0 specifies two complementary escalation pathways by which mandatory advisory authority may be activated, distinguished by whether the initiating evidence is provided by the RCU's own sensor fusion or by the traffic management authority acting on independently verified situational awareness.

\textbf{Pathway 1 --- Autonomous escalation (primarily in-coverage Category C).} When the Category C hazard occurs within the RCU's sensor coverage zone, mandatory advisory status may be activated autonomously. Three conditions must be simultaneously satisfied. First, the RCU's sensor fusion must verify a Category C hazard event --- an environmental condition producing lane blockage or triggering cascading secondary incident risk --- through its own observation layer, independent of any UE-originated report. Sensor-fusion verification is required because escalated mandatory advisories carry higher authority than standard Mode 0 messages and must not be triggered by unverified UE reports alone. Second, the approaching vehicle density within the hazard zone's stop-distance horizon must exceed a configurable threshold indicating that individual advisory compliance cannot be guaranteed within the available time window. Third, the RCU must have received regulatory pre-authorization from its governing traffic management authority to activate mandatory advisory status for the verified hazard category --- an authorization granted at deployment configuration time, not negotiated in the moment of the emergency.

When all three conditions are satisfied, the RCU broadcasts a mandatory stop advisory within the M0 resource pool using a designated high-priority message format distinguishable from standard advisory messages. The mandatory stop advisory specifies the affected lane or carriageway segment, the minimum stopping horizon, and the duration of the mandatory status. Vehicle UEs receiving the mandatory stop advisory are expected --- by design of their V2X application layer --- to treat it as a high-confidence input to their safety planning systems, analogous to how a red traffic signal is treated: not as a suggestion but as a binding instruction within the traffic management regulatory framework. The mandatory advisory is time-bounded: it remains active only for the duration verified by the RCU's sensor fusion as the active Category C condition. When the physical hazard is resolved --- by emergency services, environmental change, or lane clearance --- the RCU broadcasts a mandatory status cancellation message, and Mode 0 returns to its advisory default. The entire escalation episode is logged in the RCU's tamper-resistant record for post-incident review by the traffic management authority.

\textbf{Pathway 2 --- Commanded escalation (primarily out-of-coverage, or when in-coverage sensors are impaired).} A structurally distinct and equally important scenario arises when the Category C event occurs outside any RCU's sensor coverage zone. Road surface damage in a blind-spot approach corridor, sudden dense fog at an unmonitored tunnel exit, or flash freezing on an elevated interchange between two RCU deployment points are canonical examples. In this case, Condition 1 of the autonomous pathway cannot be satisfied: the nearest capable RCU at location Y has no sensor visibility of the incident at location X, and cannot observe cascade effects even if they have begun --- vehicles in Y's zone may still be traveling normally toward X with no deceleration or collision detectable from Y's vantage point. The autonomous pathway is architecturally inaccessible.

However, traffic management authorities routinely possess situational awareness through channels that do not depend on RCU sensor coverage: traffic monitoring facilities without computing units --- cameras, loop detectors, and sensors deployed at or near the incident location --- provide visual or behavioral evidence of the Category C event; survivor or witness reports communicated through emergency channels confirm the hazard; and coordination with emergency services provides authoritative ground truth. When such evidence establishes with sufficient certainty that a Category C event has occurred at location X, a qualified traffic management authority representative may issue a direct real-time override command to the nearest capable RCU at location Y, authorizing mandatory advisory activation for vehicle approaches to location X through Y's zone. This commanded escalation replaces Condition 1 with human judgment --- the authorizing representative serves as the verification authority in place of the RCU's sensor fusion. Conditions 2 and 3 remain active. Condition 2, the vehicle density threshold, is evaluated on what the RCU at Y can observe: the vehicles in Y's zone that are approaching X within the stop-distance horizon, not the traffic at X, which Y cannot see. Condition 3 requires pre-authorization for commanded escalation as a distinct mode.

The accountability structure of commanded escalation differs from autonomous escalation in one important respect: the authorizing representative's identity and credentials are logged in the RCU's tamper-resistant episode record alongside the standard escalation log. This creates human accountability for the mandatory command that is absent from the sensor-verified autonomous pathway, and provides the post-incident audit trail required for regulatory review. Traffic management authorities must include the commanded escalation authorization pathway in their deployment configuration, specifying which personnel roles are permitted to issue override commands and under what evidentiary threshold.

Commanded escalation separates observation, judgment and broadcast in a way that resembles the layer separation of Section~\ref{sec:8.7}, though by a different mechanism. Traffic monitoring facilities at uncovered locations --- cameras and sensors without local computing capability --- act as distributed observation nodes whose data is interpreted by human operators rather than by an RCU evaluation layer. The observation then takes place at location X, the judgment is made by the human traffic authority, and the mandatory broadcast is issued by the RCU at location Y under human direction. The RCU's mandatory broadcast capability therefore extends beyond its own sensor coverage when activated by human command, which matters for deployment economics: a traffic authority needs RCU coverage at every location where a mandatory stop advisory must be broadcast, not at every hazard-prone location.

It is essential to be precise about what the two-pathway escalation framework does and does not require. Neither pathway grants the RCU autonomous authority without institutional backing --- Pathway 1 requires pre-authorization and sensor verification; Pathway 2 requires pre-authorization and a real-time human override command. Neither pathway requires modifications to the vehicle UE's physical layer or resource allocation behavior. Neither pathway conflicts with Mode 3/1 or Mode 4/2 operations --- the mandatory advisory occupies the M0 resource pool and is distinct from all base station scheduling operations.

The two-pathway escalation framework clarifies a further dimension of Mode 0's non-substitutability. An architecture that is only ever advisory can in principle be approximated by UE-side algorithms that achieve high compliance rates. Mandatory escalation, autonomous or commanded, cannot be approximated in this way: no UE has the institutional authority to issue a mandatory stop command to other UEs (the NR-V2X Mode 2 peer equality principle explicitly forecloses this), and no UE can receive a real-time override command from a traffic management authority and execute it as a zone-wide mandatory broadcast. The capability is available only to an entity that is categorically distinct from UEs and institutionally positioned within the traffic authority domain, which is the entity Mode 0 proposes for standardization.

\subsection{Relationship to Existing Modes and Generational Independence}\label{sec:5.3}

Mode 0 is complementary to, not competitive with, the existing mode framework. A vehicle within RCU zone coverage may simultaneously use the Uu cellular interface for M1-class data services, operate in NR-V2X Mode 2 for autonomous V2V safety messaging within the M0 resource pool, and receive Mode 0 sub-zone advisories from the RCU that inform its Mode 2 resource selection. The Uu interface governs the vehicle's relationship with the cellular network for M1-class services in this scenario. NR-V2X Mode 2 governs the vehicle's autonomous V2V communication decisions. Mode 0 provides the advisory layer that improves the interference geometry of those NR-V2X Mode 2 decisions by supplying global zone information that Mode 2's local sensing cannot access. This scenario is a scoping choice rather than an architectural limit: the orthogonality established in Section~\ref{sec:4.2} means M0-class traffic carried over Uu remains equally possible, and extending Mode 0's advisory principle to that case is architecturally consistent, though the specific mechanism differs substantially from the PC5 case addressed here and is left for future work.

Mode 0's generational independence is a defining property. Mode 0 is defined at the architectural level rather than the radio access technology level. The RCU's advisory function --- zone partitioning based on kinematic data, M0/M1 resource separation, warning coverage validation, and Category C emergency escalation --- does not depend on whether the underlying air interface is LTE, NR, or a future 6G standard. The advisory architecture is invariant. This is the direct consequence of Mode 0 being defined at the entity and function level rather than the protocol level.

\subsection{Data Locality and Privacy Properties}\label{sec:5.4}

An architectural property of Mode 0 that deserves explicit treatment is the data locality guarantee it produces --- superior to both traditional cloud-dependent MEC architectures and federated learning approaches, without requiring any additional privacy mechanism.

In a cloud-dependent MEC architecture, vehicle kinematic data is aggregated at a remote server for scheduling computation. Even when the server is an edge cloud rather than a centralized cloud, the data necessarily travels beyond the traffic node boundary, crossing operator infrastructure and potentially multiple administrative domains. Trajectory data is deeply personal, enabling inference of home locations, daily routines, and behavioral patterns. Transmitting it to operator infrastructure subjects it to operator data governance policies rather than traffic authority governance policies.

Federated learning offers a partial remedy by keeping raw data on each device and sharing only model gradients. However, gradient inversion attacks can reconstruct training data from shared gradients, and the federated approach still requires communication of model parameters across the network at each training round.

The Mode 0 RCU architecture avoids both problems through geographic data containment. The RCU's evaluation layer processes only the data of vehicles currently within its bounded traffic zone, and that data circulates only within the RCU's local computational environment. A vehicle's position and kinematic state are visible to the RCU only for the duration of its zone traversal --- typically seconds to minutes --- and are not transmitted beyond the zone boundary except in aggregated, anonymized, or event-triggered form as specified by the traffic authority's data governance policy. The zone boundary is not merely a computational scope --- it is a data governance boundary enforced by the physical architecture.

This data locality property is a direct consequence of the institutional boundary argument in Section~\ref{sec:4.1}: because the RCU is owned by the traffic authority rather than the network operator, its data governance is determined by transportation law rather than telecommunications law. In jurisdictions with strong data protection requirements for personal location data --- including the European Union's GDPR framework and China's Personal Information Protection Law --- the Mode 0 architecture's zone-local data containment is structurally aligned with privacy-by-design principles in a way that cloud-dependent alternatives cannot match without additional technical controls. Full regulatory compliance additionally depends on procedural elements --- consent mechanisms, retention policies, breach notification --- that this architectural discussion does not address. This analysis assumes the observation, communication, and evaluation layers are co-located at a fixed installation; Section~\ref{sec:8.7} addresses the distinct data-locality question that arises when the evaluation layer is physically disaggregated from the other two.

\subsection{The Priority Argument: Why Mode 0 and Not Mode 5 or Mode 2e}\label{sec:5.5}

The naming of a new mode category within 3GPP is not a purely terminological question. Mode numbering carries semantic content that reflects architectural relationships.

The first alternative --- sequential extension as Mode 5 --- would imply that Mode 0 is a peer addition to the existing set, a fifth option at the same taxonomic level. This is architecturally incorrect. Mode 0 is not a fifth option at the same taxonomic level as Modes 1 through 4. It is a meta-layer that defines the resource domain within which Modes 1 through 4 operate in safety-critical contexts. Mode 0 and Mode 4/2 are not alternatives between which a vehicle chooses --- they are different layers of the same operational scenario.

The second alternative --- subfamilial extension as Mode 2e --- would be the most technically damaging. Adding Mode 2e for the RCU would imply that the RCU is also not a different class --- that it is simply another UE with an additional scheduling advisory behavior. This misclassification would foreclose the RCU's institutional distinction from the UE, eliminate the basis for traffic-authority ownership, and reduce the M0/M1 resource separation to a behavioral convention among peers rather than an architectural guarantee enforced by infrastructure-level resource partitioning. It would also eliminate the Category C mandatory escalation capability entirely, because that capability depends on the RCU's categorical distinction from UEs.

Mode 0 is proposed as numerically prior to the existing modes to reflect a specific architectural claim: within the M0 safety-critical traffic domain, the RCU's advisory layer takes precedence over the scheduling decisions of Modes 1 through 4 in terms of informational authority. The RCU's zone-level global awareness is informationally superior to any individual vehicle's local sensing, and to any base station's aggregate optimization, for the specific purpose of M0 safety scheduling within a bounded traffic zone.

The university analogy makes this layered relationship intuitive, and can be stated in its complete form. Mode 4/2 is self-directed study: the student selects resources autonomously based on local knowledge. Mode 3/1 is the general education curriculum: institutional scheduling authority allocates resources across all students without domain specialization. Mode 0 is the domain-specific core curriculum: specialized authority allocates resources for the specific discipline --- traffic safety --- with expertise and global visibility that neither self-directed study nor general education provides. Core curriculum takes precedence over general education in the professional domain, and both coexist with self-directed study outside the curriculum structure. The numeral zero reflects this foundational precedence.

Mode 0's generational independence --- established in Section~\ref{sec:5.3} --- reinforces this naming logic: it is not a peer addition within any one generation's mode set, which is precisely why it cannot be named as a sequential extension (Mode 5) or a generational variant (Mode 2e) of either.

\section{System Model and Proof-of-Concept Simulation Design}\label{sec:6}

\subsection{Mapping Mode 0 to the MARL Architecture}\label{sec:6.1}

The Mode 0 architecture of Section~\ref{sec:5} maps onto the Centralized Training, Distributed Execution (CTDE) paradigm of multi-agent reinforcement learning, and the mapping fixes what the simulation can and cannot test. In CTDE, a critic with access to the global state guides the training of decentralized actors; at execution the critic is removed, and each actor acts on its own local observation without inter-agent communication.

In the simulation, the actors model the vehicle UEs, and the RCU's zone-wide view enters training through the critic, which receives complete, accurate kinematic state for all vehicles --- an RCU whose Seeing and Thinking layers maintain an accurate mobility model (Section~\ref{sec:5.1}); degraded sensing would open a gap the simulation does not capture. As in Mode 0, nothing gates transmissions at execution time: the RCU's advisory function is not a runtime scheduler in the Mode 3/1 sense.

Three tiers of the Mode 0 subfamily (Section~\ref{sec:5.2}) are simulated. Mode 0a --- per-class shared actors with a scalar vehicle-identity feature --- models a fully passive fleet, in which the RCU serves every vehicle of a class with one shared policy. Mode 0c --- one actor per vehicle, with independent parameters --- models a fully active fleet. Mode 0b, the transitional hybrid, gives some vehicles their own actors while the passive vehicles of each class share one actor; all are trained jointly. Holding the critic fixed while varying the tier isolates the marginal value of per-vehicle UE agency; varying the critic isolates the value of its information.

The RCU's execution-side functions enter as configurations rather than learned components: the subfamily tier, which decides which vehicles run their own actors, and class-aware provisioning --- reserving a subchannel pool for safety traffic (Section~\ref{sec:6.3}). The program tests both sides: a critic-input ablation asks whether the critic's global information matters for training (Section~\ref{sec:7.5}), and the configurations ask what the execution-side functions deliver.

\subsection{Dec-POMDP Formulation}\label{sec:6.2}

We model the Mode 0 V2X resource scheduling problem as a Decentralized Partially Observable Markov Decision Process (Dec-POMDP), defined by the tuple $(N, S, \{A_i\}, \{O_i\}, T, R, \gamma)$, where $N$ is the number of vehicle agents, S is the global state space, $A_i$ and $O_i$ are the action and observation spaces for vehicle $i$, T is the state transition function, R is the reward function, and $\gamma$ is the discount factor (Section~\ref{sec:6.5.6}).

\textbf{Global state.} The global state $s \in S$ is the information available to the RCU --- equivalently, to a critic with global access during training. It is the concatenation of every vehicle's full local observation vector (as defined below) with the per-subchannel occupancy counts --- the number of vehicles transmitting on each of the $M$ subchannels in the preceding step. The global state dimension is $N \cdot \mathrm{obs\_dim} + M$.

\textbf{Local observations.} The local observation $o_i \in O_i$ for vehicle $i$ comprises the vehicle's own two-dimensional position (x, y), in kilometers, and its own speed, normalized by the maximum highway speed (33.3 m/s). These are followed by M per-subchannel exponential-moving-average channel-quality estimates ($\alpha = 0.3$), each computed by the vehicle from the fraction of its own recent packets successfully delivered on that subchannel --- a per-vehicle private signal that avoids the herding instability a shared occupancy map would produce. Three trailing features encode the vehicle's own last-step packet delivery ratio (\texttt{last\_pdr}), a class identity bit (\texttt{vehicle\_class}: 0 for M0, 1 for M1), and a normalized agent-identity index (\texttt{vehicle\_id\_norm} $\in [0,1]$), which gives a shared policy the means to break intra-class symmetry (Section~\ref{sec:6.5.2}). The observation vector has dimension $\mathrm{obs\_dim} = 3 + M + 3$.

\textbf{Actions.} Each vehicle independently selects a joint resource allocation action $a_i \in A_i$ at each generation interval, comprising a subchannel index$^1$ $ch_i \in \{0, \dots, M-1\}$ and a discrete transmit power level $p_i \in \{-10, 0, 10, 16, 23\}$ dBm --- a modeling choice, capped at the 23 dBm maximum of UE power class 3. The action is encoded as $a_i = ch_i \cdot |P| + \mathrm{idx}(p_i)$, yielding a joint action space of size $M \cdot |P|$ per vehicle.

$^1$ Throughout this paper, ``subchannel'' refers to the NR-V2X sidelink sub-channel as defined in 3GPP \mbox{TS~38.331~\S6.3.5}~\cite{ref30} --- a set of consecutive Physical Resource Blocks (PRBs) within the sidelink resource pool, sized to accommodate a complete M0 safety message payload. This is distinct from a single PRB, which is the minimum addressable resource unit in NR. The use of ``subchannel'' as the scheduling granularity reflects the Mode 0 advisory's zone-partitioning logic, which operates at the sub-channel level rather than the individual PRB level.

\textbf{Temporal granularity and traffic saturation.} Each simulation step corresponds to one safety-message generation interval of 100 ms --- the 10 Hz CAM/BSM period of ETSI EN 302 637-2~\cite{ref2} --- implemented as a 0.1 s mobility step of the Simulation of Urban MObility (SUMO) simulator~\cite{ref31}. This granularity is distinct from the sub-millisecond slot-level TTI of the NR physical layer: resource selection in this model occurs once per generation interval, not once per physical-layer slot. Every vehicle, M0-class and M1-class alike, transmits exactly one fixed-size packet per generation interval; traffic is therefore saturated at one transmission per interval, with no queueing, no retransmission, and no packet buffering across steps. Per-step delivery is scored as the delivery probability derived from the signal-to-interference-plus-noise ratio (SINR) under the mapping of Section~\ref{sec:6.4}.

\textbf{State transition.} Vehicle positions and velocities evolve according to the SUMO microscopic traffic simulator~\cite{ref31}, using its default Krauss car-following model with driver imperfection $\sigma = 0.5$; lane changes follow SUMO's default LC2013 model. The channel state transitions according to the 3GPP TR 37.885~\cite{ref32} model described in Section~\ref{sec:6.4}.

\subsection{M0/M1 Traffic Model}\label{sec:6.3}

Following the demand separation architecture of Section~\ref{sec:4.2}, the vehicle population is divided into two traffic classes. For simplicity, each simulated vehicle generates exclusively M0 or exclusively M1 traffic, although the M0/M1 classification applies to individual messages (Section~\ref{sec:4.2}) and a real vehicle generates both; within-vehicle multiplexing is left for future work.

\textbf{M0-class vehicles} generate periodic safety messages (CAM/BSM) at 10 Hz, consistent with ETSI EN 302 637-2~\cite{ref2}, each with a fixed payload of 300 bytes. M0-class vehicles are evaluated on packet delivery ratio (PDR) against the requirement PDR $\ge 0.95$, at two levels --- the mean and WIRC, defined below. The 20 ms latency bound comes from 3GPP TR 22.885~\cite{ref1}, which sets it for pre-crash sensing warnings; the 95\% level is our own choice, because TR 22.885 states no percentile. The derivation given in this section shows how the two combine. The periodic safety messages simulated here are not themselves pre-crash warnings; applying the pre-crash bound to all M0 traffic is a design choice, made because it is the tightest latency figure of the safety applications that TR 22.885 treats. The PDR level is not set by~\cite{ref1}, which asks only for ``high reliability'' without a number; it is consistent with the service-level requirements in~\cite{ref22}.

\textbf{M1-class vehicles} generate non-safety traffic representing infotainment and general data services, modeled as a constant-rate stream of one fixed-size packet per generation interval --- identical to M0 at the channel-access level, with the classes differentiated by reward structure (Section~\ref{sec:6.6}) and, under a safety partition, by pool assignment. M1-class vehicles are evaluated on mean PDR as a soft optimization target.

\textbf{Latency-tail metric.} Each message receives exactly one delivery opportunity in its generation interval, and a failed message is not retransmitted: its information can be superseded only by the next periodic message, a full generation interval (100 ms) later --- five times the 20 ms pre-crash budget. Provided the single transmission opportunity falls within 20 ms of the message's generation, the probability that the information arrives within the budget therefore equals the probability that its message is delivered: P(information within 20 ms) = per-message PDR; if the opportunity could fall later, PDR would only bound that probability from above. Reading the 20 ms bound of~\cite{ref1} at the 95th percentile --- our choice, since~\cite{ref1} states no percentile --- thus gives PDR $\ge 0.95$ on the mean: at least 95\% of messages deliver their information within 20 ms. We also apply it at the worst interval --- a stricter criterion of our own --- formalized as the Worst-Interval Reliability Criterion (WIRC): \texttt{m0\_pdr\_p05\_intra}, the 5th-percentile value of the per-interval M0 packet delivery ratio within each episode, averaged over evaluation episodes. WIRC is the metric this paper uses for the tail of the latency requirement. Its limitation is stated plainly: no end-to-end latency components --- queueing, processing, scheduling delay, or retransmission --- are modeled, so the metric bounds per-interval delivery success rather than measuring latency itself. Because WIRC averages over the safety vehicles present, it rises with their number even when protection does not improve, so WIRC values are never compared across densities.

\textbf{Population and load.} The fleet is 50\% M0-class and 50\% M1-class (\texttt{m0\_ratio} = 0.5), so $N$ vehicles include $m_0 = \lfloor N/2 \rfloor$ safety vehicles; this count ratio is independent of the data-volume asymmetry of Fig.~\ref{fig:2}. Two load ratios organize the results. The safety load $\lambda_0 = m_0 / M$ compares safety demand with subchannel supply. Below $\lambda_0 = 1$, at least one subchannel is free of safety traffic, so a placement that keeps every safety vehicle clear of M1 traffic always exists --- the M1 vehicles can share the free subchannels among themselves --- although a learner need not choose it (Sections~\ref{sec:7.4},~\ref{sec:7.6}). At $\lambda_0 = 1$ --- the \emph{critical safety load} --- each safety vehicle can still hold its own subchannel, but no placement keeps every safety vehicle clear of M1 traffic: every M1 transmission must share a subchannel with a safety vehicle. Above it, safety vehicles must share with each other. Inside a partition reserving $M_{m_0}$ subchannels for safety traffic, the pool utilization $\rho_{\text{pool}} = m_0 / M_{m_0}$ plays the same role: at $\rho_{\text{pool}} \le 1$ a collision-free safety assignment exists, while at $\rho_{\text{pool}} > 1$ the pigeonhole principle forbids one.

\subsection{Channel Model}\label{sec:6.4}

The physical layer channel model implements the 3GPP TR 37.885~\cite{ref32} V2V highway path loss model. Throughout this section, ``subchannel'' refers to the NR-V2X sidelink sub-channel definition noted in Section~\ref{sec:6.2}.

\textbf{Path loss.} Line-of-sight (LOS) path loss follows Table 6.2.1-1 of TR 37.885:

\begin{equation}\label{eq:pl}
\mathrm{PL}(d) = 32.4 + 20\log_{10}(f_c) + 20\log_{10}(d)\ [\mathrm{dB}],
\end{equation}

with $f_c = 5.9$ GHz and d clamped to 1 meter minimum.

\textbf{Shadowing.} Log-normal shadow fading with $\sigma = 3$ dB, consistent with the TR 37.885 highway LOS specification, applied to the intended link and drawn afresh in every generation interval; interfering links carry path loss and fast fading but no shadowing.

\textbf{Receiver.} Each transmission is scored at one receiver: the transmitter's nearest other vehicle in that interval (vehicle j below). If that receiver transmits on the same subchannel in the same interval, its own transmission enters the SINR as interference at the 1-meter distance floor, so the message is lost; transmissions on other subchannels do not block reception.

\textbf{Fast fading.} Rayleigh envelope fading, equivalent to Nakagami-m fading with m = 1.

\textbf{SINR computation.}

\begin{equation}\label{eq:sinr}
\mathrm{SINR}_{ij} = \frac{P_i\,G_{ij}}{\displaystyle\sum_{\substack{k \ne i\\ ch_k = ch_i}} P_k\,G_{kj} + N_0},
\end{equation}

where $N_0 = -$114 dBm, a fixed modeling constant equal to thermal noise ($-$174 dBm/Hz) over 1 MHz without a receiver noise figure.

\textbf{PDR mapping.} Instantaneous SINR is converted to a per-interval delivery probability through a logistic approximation to a block-error-rate (BLER) curve, BLER = 1 / (1 + exp(1.8 $\cdot$ ($\mathrm{SINR}^{\mathrm{dB}}$ $-$ 1.5))), with parameters intended to represent quadrature phase-shift keying (QPSK) at code rate 1/2. A 300-byte message fits in one NR code block (at most 8,448 bits), so PDR = 1 $-$ BLER: delivery is 50\% at 1.5 dB and reaches 0.95 at about 3.1 dB. The parameters are a modeling assumption, not fitted to link-level simulation; absolute delivery levels --- and with them the requirement verdicts of Table~\ref{tab:2} --- depend on them, and every scheme is evaluated under the same mapping (Section~\ref{sec:7.9}).

\subsection{MAPPO Algorithm}\label{sec:6.5}

We employ Multi-Agent Proximal Policy Optimization (MAPPO)~\cite{ref14}, implementing the CTDE architecture whose correspondence to Mode 0 was established in Section~\ref{sec:6.1}.

\subsubsection{Algorithm Selection Rationale}\label{sec:6.5.1}

MARL-based approaches have been established for dynamic spectrum and subchannel allocation in V2X~\cite{ref13,ref33}. MAPPO suits the Mode 0 scenario: the RCU's zone is physically bounded, so its global state --- up to about 100 vehicles within 300 meters --- remains tractable for one multilayer perceptron (MLP) critic. MAPPO is also a strong and simple default for cooperative multi-agent problems~\cite{ref14}, without the value-decomposition machinery of methods such as QMIX and QTRAN. MAPPO needs no inter-vehicle communication in training or execution, and since training uses only zone-local data, it inherits the data containment of Section~\ref{sec:5.4}. Whether the critic needs the global state at all is tested directly (Section~\ref{sec:7.5}).

\subsubsection{Actor Networks}\label{sec:6.5.2}

In the Mode 0a configuration, each traffic class is served by a dedicated shared actor network --- one MLP with two hidden layers of 128 units each (ReLU activations; 11--128--128--25, 21,273 parameters at $M = 5$) shared across all m0 M0 vehicles, and a separate network of identical architecture shared across all M1 vehicles. The \texttt{vehicle\_id\_norm} observation feature lets the shared policy condition on vehicle identity without introducing cross-class parameter coupling. Both actors output logits over the joint action space $A_i$; the policy $\pi_\theta(a_i \mid o_i)$ is a categorical distribution over actions. In the Mode 0c configuration, each vehicle operates its own actor network with identical architecture but separate parameters --- $N$ distinct networks rather than two shared-class networks. In Mode 0b, active vehicles have their own networks while the passive vehicles of each class share one, trained on their pooled experience.

The symmetric-Nash floor is the analytical baseline for coordination. For $n$ vehicles selecting uniformly among $M$ subchannels, the probability that a given vehicle shares its subchannel is $P_{\text{floor}} = 1 - ((M-1)/M)^{n-1}$; within a class, $n$ is that class's vehicle count (five safety vehicles on five subchannels: $P_{\text{floor}} = 0.590$). With exchangeable channels, uniform randomization is a best response to uniform randomization, and no non-uniform symmetric profile is an equilibrium, so $P_{\text{floor}}$ is the collision probability of the unique symmetric equilibrium of the idealized single-shot game --- not a bound over all policies, since asymmetric equilibria with distinct assignments exist whenever the pool permits them. A shared network with symmetric reward can reach one only by conditioning on the identity feature~\cite{ref15}; whether it does depends on the size of the class group it serves (Section~\ref{sec:7.4}). Per-vehicle actors are asymmetric from initialization.

\subsubsection{Critic Networks}\label{sec:6.5.3}

Three critics are compared (Section~\ref{sec:7.5}). The original \emph{team-value} critic outputs one scalar $V_\psi(s)$ from the global state $s \in \mathbb{R}^{N \cdot \mathrm{obs\_dim} + M}$, shared by all vehicles; a \emph{local} critic outputs a per-vehicle value $V_\psi(o_i)$ from the vehicle's own observation; an \emph{agent-specific} critic outputs a per-vehicle value $V_\psi(s \oplus o_i)$ from both. Agent-specific against local isolates the critic's information; agent-specific against team value isolates its structure. All three are MLPs with two hidden layers of 256 units and LayerNorm after the first (96,257, 69,633 and 99,073 parameters at $N = 10$, $M = 5$), used only in training. Mode 0c is reported with the team-value critic, on which its results do not depend; Mode 0a and Mode 0b with the agent-specific critic, the strongest baseline for shared actors (Section~\ref{sec:7.5}).

\subsubsection{Generalized Advantage Estimation}\label{sec:6.5.4}

Advantages are computed per vehicle, along each vehicle's own trajectory, with $\lambda = 0.95$:

\begin{equation}\label{eq:gae}
\begin{split}
\hat{A}_t^i &= \sum_{l=0}^{\infty} (\gamma\lambda)^l\,\delta_{t+l}^i, \\
\delta_t^i &= r_t^i + \gamma V_{t+1}^i - V_t^i,
\end{split}
\end{equation}

where $V_t^i$ is the critic's value for vehicle $i$ at step $t$ --- the team value, or the vehicle's own value under a per-vehicle critic. Summing over the transitions of all vehicles as a single sequence would mix one vehicle's rewards and values into another's advantages.

\subsubsection{PPO-Clip Objective}\label{sec:6.5.5}

\begin{equation}\label{eq:ppo}
\begin{split}
L^{\mathrm{CLIP}}(\theta) = \mathbb{E}_t\bigl[\min\bigl(\rho_t\hat{A}_t,\ \\
& \hspace{-2.0cm}\mathrm{clip}(\rho_t, 1-\varepsilon, 1+\varepsilon)\,\hat{A}_t\bigr)\bigr]
\end{split}
\end{equation}

where $\rho_t = \pi_\theta(a_t \mid o_t) / \pi_{\theta_{\mathrm{old}}}(a_t \mid o_t)$ and $\varepsilon = 0.2$. An entropy bonus is added to the actor objective with coefficient \texttt{ent\_c} annealed linearly from 0.01 at training start to 0.001 at training end across 5,000 episodes. Evaluation always uses the final entropy coefficient.

\subsubsection{Training Hyperparameters}\label{sec:6.5.6}

The discount factor is $\gamma = 0.5$. The objective is delivery in each generation interval, so returns need not reach far: with $\lambda = 0.95$, $\gamma = 0.99$ would make each advantage a discounted sum over about 17 future steps ($\gamma\lambda \approx 0.94$) and inflate its variance by a factor of about 8.7, against 1.29 for $\gamma = 0.5$ ($\gamma\lambda = 0.475$). Actor and critic learning rates are both $3 \times 10^{-4}$, cosine-annealed to a floor of $10^{-5}$ over 5,000 training episodes; gradient norms are clipped at 0.5. Each episode lasts 200 generation intervals (20 s of simulated traffic) and forms one training batch, with 4 PPO update epochs per episode. Training runs for 5,000 episodes, followed by 100 evaluation episodes at terminal-entropy stochastic action selection.

\subsection{Reward Function Design}\label{sec:6.6}

\textbf{M0 vehicle reward:}

\begin{equation}\label{eq:rm0}
\begin{split}
r_i^{\mathrm{M0}} = {}& \alpha\,\mathrm{PDR}_i + \beta\,\mathrm{clip}\!\left(\tfrac{\mathrm{SINR}_i^{\mathrm{dB}}}{20},0,1\right) \\
& + \gamma_{\mathrm{team}}\,\mathrm{mean}\bigl(\{\mathrm{PDR}_j : j \in \mathrm{M0}\}\bigr),
\end{split}
\end{equation}

with $\alpha = 1.0$, $\beta = 0.3$, $\gamma_{\text{team}} = 0.5$.

\textbf{M1 vehicle reward:}

\begin{equation}\label{eq:rm1}
\begin{split}
r_i^{\mathrm{M1}} = {}& \delta\,\mathrm{clip}\!\left(\tfrac{\mathrm{SINR}_i^{\mathrm{dB}}}{20},0,1\right) \\
& - \eta\,\bigl(1 - \mathrm{mean}(\mathrm{PDR}_{\mathrm{M0}})\bigr),
\end{split}
\end{equation}

with $\delta = 0.3$ and $\eta = 0.3$. The $\eta$ term couples M1's reward to M0 outcomes: when M0 PDR falls, M1 agents are penalized, providing gradient pressure for M1 to avoid resource decisions that displace M0 even when no partition is enforced.

The asymmetry between M0 and M1 reward magnitudes is deliberate, so that the learned policy prioritizes M0 delivery even in a shared pool. The reward is nonetheless a weighted sum, and a weighted sum does not enforce the paper's safety-first premise lexicographically: a learner may give up safety delivery to keep M1 vehicles apart, as Sections~\ref{sec:7.4} and~\ref{sec:7.6} show. Sensitivity to the class weight is tested by raising $\delta$ from 0.3 to 1.0 (Section~\ref{sec:7.5}).

\subsection{Simulation Environment and Experimental Design}\label{sec:6.7}

\textbf{Co-simulation architecture.} SUMO provides vehicle mobility simulation via the TraCI interface. A standalone C++ channel simulation binary connected through a ZeroMQ REQ/REP socket implements the TR 37.885 channel model. The Python/PyTorch training environment orchestrates both simulators.

\textbf{Highway scenario.} A 3 km straight highway segment with 2 lanes in a single direction of travel; vehicle parameters: acceleration 2.6 m/s$^2$, deceleration 4.5 m/s$^2$, maximum speed 33.3 m/s, vehicle length 5.0 m, minimum gap 2.5 m, driver imperfection $\sigma = 0.5$ (Krauss). The number of transmitting vehicles ranges from $N = 4$ to $N = 20$.

\textbf{Density control.} SUMO injects vehicles at 1,200 vehicles per hour in every run. At reset, the simulation advances until $N$ vehicles are on the road; these first $N$ vehicles are the agents, and vehicles entering later move in the traffic stream without transmitting. Classes are assigned in blocks by order of entry: the first m0 vehicles are safety vehicles and the rest M1, so the safety class starts at the front of the agents' platoon (overtaking may mix the order). Every scheme sees the same layout, but it shapes absolute delivery and the reuse ratio, since a safety vehicle and an M1 vehicle sharing a subchannel tend to be far apart. Because the inflow is the same for every N, a larger N lengthens the stretch of road the agents occupy rather than packing a fixed stretch more densely. Delivery levels therefore do not scale with N as they would in a fixed-length zone, and coordination is compared across $N$ through within-class collision rates rather than delivery (Section~\ref{sec:7.4}).

\textbf{Experimental program.} $M = 5$ and \texttt{m0\_ratio} = 0.5 unless stated. (i) \emph{Capability ladder}: SB-SPS, Mode 0a and Mode 0c at $N = 4$ ($\lambda_0 = 0.4$) and $N = 10$ ($\lambda_0 = 1$), with two ablations of Mode 0c --- Arm S, which fixes power at 23 dBm while subchannel choice remains learned, at both loads, and Arm P, which draws each subchannel uniformly at random every interval while power remains learned, at $N = 10$ only. (ii) \emph{Critic-input ablation}: the three critics for Mode 0c at $N = 4$ and 10 and for Mode 0a at $N = 4$; at $N = 10$, Mode 0a with the team-value and agent-specific critics. (iii) \emph{Reward weight}: Mode 0c at $N = 10$ with $\delta = 1.0$. (iv) \emph{Class-group sweep}: Mode 0a (agent-specific critic) at N $\in$ {4, 5, 6, 7, 10, 15, 20} and at $N = 4$ with M $\in$ {3, 4} --- two to ten safety vehicles per shared actor. (v) \emph{Safety partition}: Mode 0c with $M_{m_0} = 2$ at $N = 4$ ($\rho_{\text{pool}} = 1$); $M_{m_0} = 2$ and 4 at $N = 10$ ($\rho_{\text{pool}} = 2.5$, 1.25); and $M_{m_0} = 5$ at $N = 10$, $M = 7$ ($\rho_{\text{pool}} = 1$, $\lambda_0 \approx 0.71$), each against the shared pool at the same N and M. (vi) \emph{Mixed fleet} (Mode 0b) at $N = 10$: two of five safety and three of five M1 vehicles passive, each passive group sharing one class actor. Each configuration has three seeds, except Mode 0c with the team-value critic at $N = 4$ and 10, with five; the class-group sweep reuses the Mode 0a agent-specific runs at $N = 4$ and 10, and each partition is compared with the shared-pool runs at the same N and M --- 88 training runs in all, listed by configuration in the Supplementary Material.

\textbf{SB-SPS baseline.} The standardized reference is sensing-based semi-persistent scheduling (SB-SPS) for NR-V2X Mode 2, as specified in \mbox{TS~38.214}~\cite{ref34} (clause~8.1.4) and analyzed in~\cite{ref10}, adapted to one opportunity per generation interval. Each vehicle keeps its subchannel for a reselection counter drawn uniformly from 5 to 15 intervals (no resource keeping) and transmits at 23 dBm. At reselection it excludes subchannels that another vehicle reserved within the 10-interval sensing window and is heard above a Reference Signal Received Power (RSRP) threshold of $-$80 dBm, raising the threshold in 3 dB steps until at least 20\% of the pool remains, and draws uniformly among the candidates. Sensing is idealized in SB-SPS's favor (every reservation decoded, received power without measurement noise), while other idealizations favor Mode 0 (no priority-dependent thresholds, one opportunity per interval over five subchannels, no hybrid automatic repeat request (HARQ), exact positions at the RCU, no signaling cost); on balance the comparison flatters Mode 0, and the baseline is not an upper bound on the standard. Evaluation follows a 40-interval warm-up, two complete reselection cycles, with the residual warm-up effect running against SB-SPS; the threshold lies on a performance plateau. Pseudocode and the RSRP-threshold and warm-up sweeps are in the Supplementary Material.

\textbf{Seeds and realizations.} The headline cells --- Mode 0c and SB-SPS at $N = 4$ and 10 --- use five seeds; all others use three. Each seed fixes the Python, NumPy and PyTorch random-number generators. SUMO receives no seed, so every episode in every seed replays the same traffic realization, while shadowing and fading are drawn afresh each interval; generalization to unseen traffic realizations is not tested (Section~\ref{sec:7.9}). Results are means across seeds with two-sided 95\% t-intervals on the sample standard deviation; intervals crossing 0 or 1 are clipped and marked, and a cell whose seeds fall into two modes is reported split. Per-seed values are in the archived release.

\textbf{Performance metrics.} Safety class: mean PDR and WIRC (Section~\ref{sec:6.3}); per vehicle, the worst safety vehicle's delivery in each episode and the share of safety vehicle-episodes meeting 0.95; and information age (the Age of Information metric of~\cite{ref35}) --- generation intervals since a vehicle's safety information was last delivered, from delivery events drawn with each interval's delivery probability, whose \emph{peak} is each episode's longest staleness averaged over episodes. Coordination: the within-class collision rate against the floor of Section~\ref{sec:6.5.2}, and the spatial reuse ratio --- the mean separation of co-channel vehicles divided by the same with subchannels shuffled among vehicles each interval (above 1: farther apart than chance). Learning: \emph{coordination time}, the episodes until delivery is sustained at $\ge 0.90$ --- a learning measure, not a requirement. M1 class: mean PDR. Throughout, \emph{tail} refers to the worst-case end of safety delivery in three forms --- the worst interval (WIRC), the worst vehicle, and peak information age; the tail of the latency requirement itself is measured by WIRC (Section~\ref{sec:6.3}).

\textbf{Decision rules.} Comparisons stated as wins use a lexicographic rule fixed before each batch of runs: safety delivery first, where a difference counts only if the 95\% intervals do not overlap and the means differ by at least 0.005; M1 delivery only on a safety tie, with a floor of 0.01; coordination time last, where it counts only if the medians differ by more than a factor of two. Requirement claims are judged on the interval: a configuration \emph{meets} PDR $\ge 0.95$ only if its whole interval lies at or above 0.95, falls \emph{short} if its whole interval lies below, and otherwise \emph{straddles} the requirement.

\section{Results and Analysis}\label{sec:7}

\subsection{Program and How Results Are Judged}\label{sec:7.1}

Section~\ref{sec:7} draws on 88 training runs, all at the full budget of 5,000 episodes and listed by configuration in the Supplementary Material, together with 10 SB-SPS evaluations and 29 per-vehicle re-evaluations of trained policies and SB-SPS configurations. Results are organized by the safety load $\lambda_0 = m_0 / M$ (Section~\ref{sec:6.3}): $N = 4$ with $M = 5$ places the zone below the critical load ($\lambda_0 = 0.4$), and $N = 10$ places it at the critical load ($\lambda_0 = 1$), the lowest load at which no placement keeps every safety vehicle clear of M1 traffic. Every value is a mean across seeds with its 95\% interval; comparisons use the lexicographic rule of Section~\ref{sec:6.7}, and the requirement PDR $\ge 0.95$ is judged on the interval, at the mean and at WIRC. Two reference values recur: the within-class floor of Section~\ref{sec:6.5.2} --- 0.590 for five safety vehicles on five subchannels --- and the requirement itself.

\subsection{Result 1 --- The Critical Safety Load: Per-Vehicle Actors Against the Standard}\label{sec:7.2}

\begin{figure*}[!t]
\centering
\includegraphics[width=\figwidew,keepaspectratio]%
                {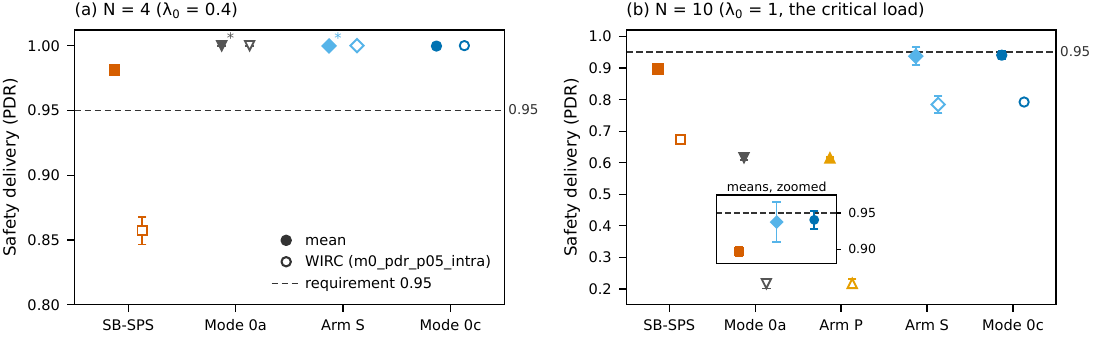}
\caption{The capability ladder at (a) $N = 4$ and (b) $N = 10$, the critical safety load. Filled markers: mean safety delivery; hollow markers: WIRC (\texttt{m0\_pdr\_p05\_intra}); bars: 95\% confidence intervals over seeds (n = 5 for SB-SPS and Mode 0c, n = 3 otherwise; * clipped at 1). Mode 0a uses the agent-specific critic, its best. Arm P draws subchannels at random with learned power; Arm S learns subchannels with power fixed at 23 dBm. The dashed line is the 0.95 requirement. Inset: the means nearest the requirement, zoomed --- Mode 0c's interval includes 0.95. The panels use different vertical scales and are not compared with each other.}
\label{fig:3}
\end{figure*}

Fig.~\ref{fig:3} presents the capability ladder at both loads, and Table~\ref{tab:2} judges each configuration against the requirement.

\textbf{Below the critical load}, every scheme delivers nearly all safety messages on average: Mode 0c 0.9997 [0.9995, 0.9998], Mode 0a 1.0000 [0.9999, 1.0000]*, SB-SPS 0.9811 [0.9800, 0.9823]. The schemes separate at the worst interval: Mode 0c and Mode 0a hold a WIRC of 1.000, while SB-SPS's is 0.8573 [0.8465, 0.8681]. SB-SPS therefore meets the requirement on the mean but falls short at the worst-interval level --- the case for reporting WIRC at all.

\textbf{At the critical load}, Mode 0c delivers 0.9405 [0.9281, 0.9528] of safety messages with a WIRC of 0.7921 [0.7815, 0.8027], and M1 delivery of 0.9629 [0.9568, 0.9689]. SB-SPS delivers 0.8967 [0.8900, 0.9034], with a WIRC of 0.6736 [0.6578, 0.6894] and M1 delivery of 0.9437 [0.9418, 0.9455]. The intervals separate on all three, and the gap is 0.044 on the mean but 0.119 on WIRC --- 2.7 times larger where the tail of the latency requirement is measured. The mechanism is visible in the collision rates. Mode 0c keeps all five safety vehicles on separate subchannels (within-class collision 0.000) and pairs each with an M1 vehicle, while SB-SPS leaves safety vehicles sharing with each other (within-class 0.126 [0.109, 0.143]), because its sensing excludes only reservations it hears above the threshold.

Shared actors fail at this load. Mode 0a delivers 0.6158 [0.6096, 0.6220], with a WIRC of 0.217, and its safety vehicles collide with each other at 0.590 [0.584, 0.596] --- the random-choice floor itself, with either critic (Section~\ref{sec:7.5}). At the critical load, a single shared policy per class performs worse than the standard. The critic is not held fixed in this headline comparison --- Mode 0c here uses the team-value critic and Mode 0a its best, the agent-specific critic --- but the gap does not depend on it: with the agent-specific critic for both, Mode 0c delivers 0.9413 [0.9113, 0.9713] (Table~\ref{tab:3}) against Mode 0a's 0.6158.

\begin{table*}[!t]
\centering
\caption{Safety delivery against the derived requirement (PDR $\geq 0.95$), judged on the 95\% interval: \textit{meets} (interval wholly at or above 0.95), \textit{short} (wholly below), \textit{straddles}. Each operating point is judged separately; WIRC is not compared across densities. $^{*}$ interval clipped at 1.}
\label{tab:requirement}
\label{tab:2}
\footnotesize
\begin{tabular}{llllll}
\toprule
Operating point & Scheme & Mean M0 PDR [95\% CI] & Verdict & WIRC [95\% CI] & Verdict \\
\midrule
$N = 4$ & Mode 0c & 0.9997 [0.9995, 0.9998] & meets & 1.0000 [1.0000, 1.0000] & meets \\
 & SB-SPS & 0.9811 [0.9800, 0.9823] & meets & 0.8573 [0.8465, 0.8681] & short \\
$N = 10$ & Mode 0c & 0.9405 [0.9281, 0.9528] & straddles & 0.7921 [0.7815, 0.8027] & short \\
 & Mode 0b & 0.9491 [0.9098, 0.9884] & straddles & 0.7988 [0.7957, 0.8019] & short \\
 & Arm S & 0.9375 [0.9097, 0.9652] & straddles & 0.7844 [0.7568, 0.8120] & short \\
 & SB-SPS & 0.8967 [0.8900, 0.9034] & short & 0.6736 [0.6578, 0.6894] & short \\
 & Mode 0a (agent-specific) & 0.6158 [0.6096, 0.6220] & short & 0.2170 [0.2017, 0.2324] & short \\
 & partition, pool 4 & 0.8785 [0.8772, 0.8799] & short & 0.6533 [0.6452, 0.6614] & short \\
$N = 10$, $M = 7$ & shared pool & 0.9732 [0.9502, 0.9963] & meets & 0.8043 [0.7876, 0.8210] & short \\
 & partition, pool 5 (RCU-sized) & 1.0000 [0.9999, 1.0000]$^{*}$ & meets & 1.0000 [1.0000, 1.0000] & meets \\
\bottomrule
\end{tabular}
\end{table*}

Against the requirement, no scheme's WIRC meets 0.95 at the critical load: every interval lies wholly below it. On the mean, SB-SPS is clearly short. Mode 0c's mean is about 0.01 below 0.95, but its interval, [0.928, 0.953], includes the requirement.

\textbf{Per vehicle}, the comparison widens rather than narrows. At the critical load the worst safety vehicle in each episode delivers 0.8251 [0.7635, 0.8868] under Mode 0c against 0.6761 [0.6639, 0.6882] under SB-SPS --- a gap of 0.149, larger than the gap in WIRC --- and safety vehicles meet 0.95 over an episode in 60\% [54, 67] of vehicle-episodes against 50\% [48, 52]. Below the critical load the shares are 100\% and 86\% [84, 87]. WIRC averages over the safety vehicles present and could, in principle, hide one vehicle's repeated failure; per vehicle, the advantage holds.

\textbf{Spatial reuse} is real and measured: vehicles sharing a subchannel under Mode 0c sit 1.419 [1.416, 1.421] times farther apart than under random assignment of the same channel occupancy, against 1.375 [1.373, 1.377] under SB-SPS's sensing. The learned policy places co-channel pairs farther apart than both chance and sensing.

\subsection{Result 2 --- What Carries the Advantage: Coordinated Subchannel Choice}\label{sec:7.3}

Mode 0c learns two things at once: which subchannel to use and what power to transmit. The two capability ablations of Fig.~\ref{fig:3}(b) separate them, against Mode 0c's matched seeds (0.9351 [0.9211, 0.9491]).

With subchannels drawn at random and power still learned (Arm P), safety delivery falls to 0.6149 [0.6126, 0.6172]. Its collision rates match uniformly random subchannels --- any collision 0.865 against 1 $-$ (4/5)$^9 = 0.866$, within-class 0.592 against 0.590 --- and its delivery matches Mode 0a's floor (0.6158) to within 0.001. Learned power control cannot compensate for uncoordinated subchannels.

With subchannels learned and power fixed at 23 dBm (Arm S), safety delivery is 0.9375 [0.9097, 0.9652], a lexicographic tie with Mode 0c on safety delivery, M1 delivery (0.9570 against 0.9610) and safety coordination time; below the critical load the two tie as well (0.9997 against 0.9997). Fixed power changes how M1 vehicles are placed --- some share among themselves --- and slows M1's coordination about threefold (a median of 1,100 episodes against 350) without changing either class's final delivery.

The advantage of per-vehicle actors therefore comes from coordinated subchannel choice. Learned power control adds nothing measurable to the delivery of either class at either load.

\subsection{Result 3 --- Where Shared Actors Fail: Class-Group Size}\label{sec:7.4}

\begin{figure*}[!t]
\centering
\includegraphics[width=\figwidew,keepaspectratio]%
                {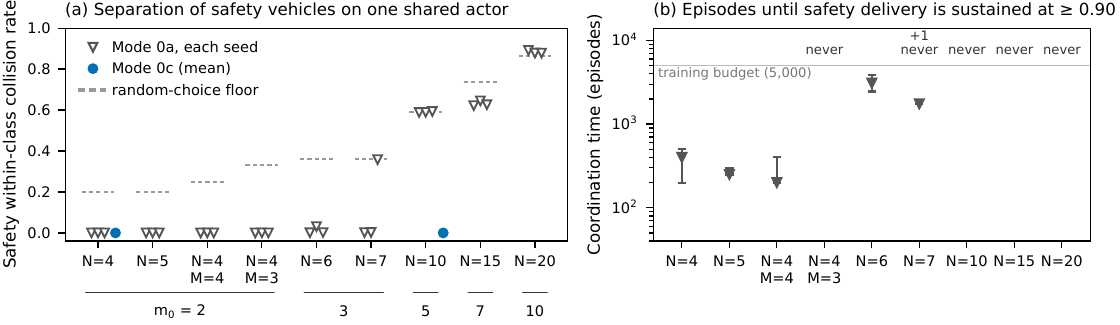}
\caption{Shared actors' separation of safety vehicles against class-group size m0 (Mode 0a, agent-specific critic; three seeds per point). (a) Within-class collision rate of the safety vehicles, each seed shown; dashed ticks give each point's random-choice floor, $1 - (1 - 1/M)^{m_0 - 1}$; diamonds show Mode 0c at $N = 4$ and 10. (b) Coordination time --- episodes until safety delivery is sustained at $\ge 0.90$ --- median and range over the seeds that reached it; ``+1 never'' marks a further seed that did not reach it within the 5,000-episode budget. At $N = 4$ with $M = 3$, four vehicles share three subchannels, and the learner pairs one safety vehicle with an M1 vehicle in every interval rather than letting the two M1 vehicles share; safety delivery stays near 0.76 and never reaches the threshold, although the safety vehicles separate completely (Section~\ref{sec:7.4}). Two safety vehicles always separate; three usually separate, slowly; five or more do not within 5,000 episodes.}
\label{fig:4}
\end{figure*}

A shared actor can separate the vehicles it serves only by conditioning on the identity feature (Section~\ref{sec:6.5.2}). Fig.~\ref{fig:4} measures how often it does as the number of safety vehicles it serves grows from two to ten.

The limit is a gradient with group size, not a sharp boundary. \textbf{Two} safety vehicles on one shared actor separate in all 12 seeds across four configurations ($N = 4$ and 5; $N = 4$ with $M = 3$ and 4), within 200--500 episodes. \textbf{Three} separate in five of six seeds ($N = 6$: three of three; $N = 7$: two of three), but slowly --- after 1,750--3,900 episodes --- and the remaining seed stays exactly on its floor (0.358 against 0.360). \textbf{Five or more} never separate within the budget (zero of nine seeds). At $N = 10$ all three seeds sit on the floor (0.588--0.593 against 0.590), under either critic; at $N = 15$ they show partial structure, 0.09--0.12 below the floor of 0.738, without separating; at $N = 20$ they lie on or slightly above the floor of 0.866. Mode 0c's safety vehicles separate completely at both $N = 4$ and $N = 10$.

Per-vehicle actors also coordinate faster where both succeed. Below the critical load, Mode 0c brings safety delivery to a sustained 0.90 four to eight times sooner than Mode 0a under every critic --- a median of 50--100 episodes against 300--400 (Table~\ref{tab:3}).

Two cautions apply to Fig.~\ref{fig:4}. Coordination time also reflects placement, not only separation: at $M = 3$, four vehicles share three subchannels, so one pair must share in every interval. The two M1 vehicles could share a subchannel and leave both safety vehicles clear; instead the learner keeps the M1 vehicles apart and pairs one of them with a safety vehicle in every interval (safety collision 0.500), so safety delivery stays at 0.756 [0.753, 0.760] although the safety vehicles separate completely --- the same placement choice as in Section~\ref{sec:7.6}. And delivery levels are not compared across $N$ (Section~\ref{sec:6.7}), which is why Fig.~\ref{fig:4} plots within-class collision. One seed at $N = 6$ separated its safety vehicles while its M1 class collapsed (0.4965), so M1 at that point is reported by seed.

The critical load is therefore exactly where shared actors give out. With five safety vehicles per class, a single shared policy falls to the random floor, and per-vehicle actors become necessary rather than merely faster.

\subsection{Result 4 --- Training: The Critic's Information and the Reward Weight}\label{sec:7.5}

Table~\ref{tab:3} compares the three critics of Section~\ref{sec:6.5.3} and the reward-weight test.

\begin{table*}[!t]
\centering
\caption{Critic-input ablation (three seeds per cell) and the reward-weight test ($\delta$). Coordination time: episodes until delivery is sustained at $\geq 0.90$, median [range]. $^{*}$ interval clipped at 1.}
\label{tab:critic}
\label{tab:3}
\footnotesize
\begin{tabular}{lllll}
\toprule
Critic & M0 PDR [95\% CI] & M1 PDR [95\% CI] & \multicolumn{2}{c}{Coordination time (episodes)} \\ \cmidrule(lr){4-5} & & & M0 & M1 \\
\midrule
\multicolumn{5}{l}{\textit{Mode 0c, $N = 10$}} \\
\quad team value $V(\mathbf{s})$ & 0.9351 [0.9211, 0.9491] & 0.9610 [0.9557, 0.9663] & 100 [50--100] & 350 [200--650] \\
\quad local $V(o_i)$ & 0.9468 [0.9142, 0.9793] & 0.9621 [0.9531, 0.9711] & 50 [50--100] & 50 [50--100] \\
\quad agent-specific $V(\mathbf{s}\oplus o_i)$ & 0.9413 [0.9113, 0.9713] & 0.9645 [0.9576, 0.9715] & 50 [50--450] & 100 [50--100] \\
\multicolumn{5}{l}{\textit{Mode 0c, $N = 4$}} \\
\quad team value & 0.9997 [0.9995, 0.9999] & 0.9987 [0.9971, 1.0000]$^{*}$ & 100 [100--100] & 150 [150--250] \\
\quad local & 1.0000 [0.9999, 1.0000]$^{*}$ & 0.9999 [0.9999, 0.9999] & 50 [50--100] & 50 [50--100] \\
\quad agent-specific & 1.0000 [0.9999, 1.0000]$^{*}$ & 1.0000 [1.0000, 1.0000]$^{*}$ & 50 [50--100] & 100 [50--100] \\
\multicolumn{5}{l}{\textit{Mode 0a, $N = 4$}} \\
\quad team value & 0.9999 [0.9998, 1.0000]$^{*}$ & 0.7962 [0.7932, 0.7992] & 400 [350--650] & never \\
\quad local & 1.0000 [0.9999, 1.0000]$^{*}$ & 1.0000 [0.9999, 1.0000]$^{*}$ & 300 [250--1050] & 250 [200--450] \\
\quad agent-specific & 1.0000 [0.9999, 1.0000]$^{*}$ & 1.0000 [1.0000, 1.0000]$^{*}$ & 400 [200--500] & 250 [150--300] \\
\multicolumn{5}{l}{\textit{Mode 0c, $N = 10$, $\delta = 1.0$}} \\
\quad team value & 0.9393 [0.9056, 0.9729] & 0.9629 [0.9515, 0.9743] & 100 [50--100] & 50 [50--100] \\
\bottomrule
\end{tabular}
\end{table*}

\textbf{The critic's information makes no difference.} A local critic, which sees only the vehicle's own observation, trains as well as an agent-specific critic that sees the whole zone, in every configuration: for Mode 0c at the critical load, 0.9468 [0.9142, 0.9793] against 0.9413 [0.9113, 0.9713].

\textbf{The critic's structure matters for shared actors, within limits.} For Mode 0c, the team value ties the per-vehicle critics on both classes; it only slows M1's coordination (350 episodes against 50--100). For Mode 0a below the critical load, the team value leaves the shared M1 actor unseparated: its two vehicles stay off the safety subchannels but choose at random among the three left free (within-class collision 0.336, against 1/3 at random; M1 delivery 0.7962 [0.7932, 0.7992]), while either per-vehicle critic coordinates it fully (1.0000). A per-vehicle value thus helps a shared actor separate a group of two. It does not rescue a group of five: at the critical load, Mode 0a stays on the floor with the agent-specific critic (0.6158) as with the team value (0.6161), within-class 0.590 under both. Mode 0a is therefore reported with the agent-specific critic throughout, so that its results are not handicapped by its critic.

These results locate the RCU's contribution. The CTDE mapping of Section~\ref{sec:6.1} remains a valid model of how an RCU could train vehicle policies, but the simulation shows no advantage from the critic's zone-wide view: a local critic does as well. The RCU's measured value lies instead on the execution side --- choosing and sizing class-aware provisioning (Section~\ref{sec:7.6}) and serving vehicles within the subfamily (Section~\ref{sec:7.7}) --- alongside the authority and information functions argued in Sections~\ref{sec:3}--\ref{sec:5}.

\textbf{The results do not depend on the reward's class weight.} Raising M1's weight from $\delta = 0.3$ to 1.0 leaves both classes unchanged at the critical load --- safety delivery 0.9393 [0.9056, 0.9729] against 0.9351 [0.9211, 0.9491], M1 delivery 0.9629 against 0.9610, a lexicographic tie on both and on coordination time. The learned structure is also unchanged: every safety vehicle shares with an M1 vehicle, none with another safety vehicle. At $M = 5$ every safety vehicle must share with an M1 vehicle anyway, so the weight can act only through placement and power, and neither moves. This covers one coefficient at two values and one operating point; where spectrum is spare, the learner chooses how many M1 vehicles occupy safety subchannels (Section~\ref{sec:7.6}), and the weight may matter there.

\subsection{Result 5 --- The Safety-Priority Partition}\label{sec:7.6}

\begin{figure*}[!t]
\centering
\includegraphics[width=\figwidew,keepaspectratio]%
                {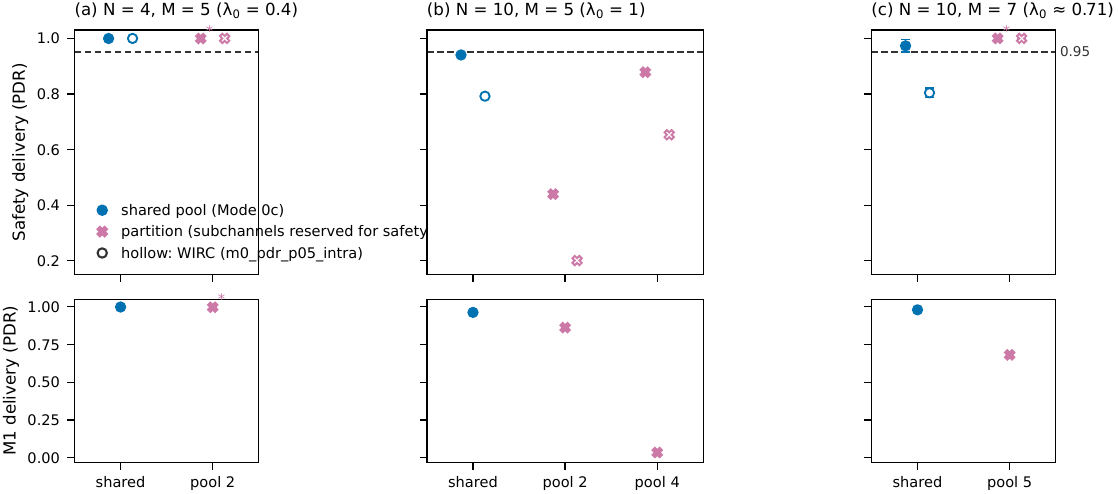}
\caption{The safety-priority partition at three operating points: (a) $N = 4$, (b) the critical load $N = 10$ with $M = 5$, (c) spare spectrum, $N = 10$ with $M = 7$. Top: safety delivery (filled: mean; hollow: WIRC (\texttt{m0\_pdr\_p05\_intra})) with the 0.95 requirement; bottom: M1 delivery. Circles: shared pool (Mode 0c); crosses: a partition reserving the indicated number of subchannels for safety traffic (pool 2 and pool 4 at the critical load; pool 5, sized to the safety demand by the RCU, at $M = 7$). Only with spare spectrum does a partition sized to the safety demand meet the requirement at both levels; at the critical load every partition loses to the shared pool.}
\label{fig:5}
\end{figure*}

A partition reserves $M_{m_0}$ subchannels for safety traffic and confines M1 to the rest. Fig.~\ref{fig:5} tests it at three operating points.

\textbf{Below the critical load}, a partition changes nothing: at $N = 4$ with two of five subchannels reserved, safety delivery is 0.9997 [0.9990, 1.0000] against 0.9997 shared, and M1 delivery 0.9968 against 0.9984.

\textbf{With spare spectrum} ($N = 10$, $M = 7$), a partition sized to the safety demand --- five subchannels for five safety vehicles --- delivers every safety message, mean 1.0000 [0.9999, 1.0000] and WIRC 1.000. The shared pool delivers 0.9732 [0.9502, 0.9963], with a WIRC of 0.8043 [0.7876, 0.8210]. The partition's cost falls on M1, whose delivery drops from 0.9800 to 0.6811 with five M1 vehicles confined to two subchannels.

The shared pool's mechanism shows why the partition is needed. The learned policy keeps every M1 vehicle away from every other: two take the free subchannels, and three sit on safety subchannels, so that three of the five safety vehicles share in every interval (safety collision rate 0.600 in every seed). It gives up 0.027 of safety delivery to gain 0.299 of M1 delivery --- a net gain under the 1.0 : 0.3 weighted reward. A weighted-sum reward does not enforce the paper's safety-first premise; a partition sized to the safety demand does. It is the only configuration at $N = 10$ that meets the requirement at both levels; per vehicle, every safety vehicle delivers at least 0.95 in every episode, and peak staleness falls to 1.05 generation intervals.

\textbf{At the critical load}, every partition loses to the shared pool. Reserving four of five subchannels leaves five safety vehicles on four, so the best split forces one pair to share (within-class collision 0.400 in every seed), and safety delivery falls to 0.8785 [0.8772, 0.8799] against 0.9405 shared, while all five M1 vehicles crowd onto the one remaining subchannel (M1 0.035). Reserving two leaves every safety vehicle sharing (within-class 1.000; delivery 0.4396). Reserving all five would leave M1 no subchannel at all.

The sizing rule therefore carries a condition. Where spectrum is spare (m0 $<$ M, $\lambda_0 < 1$), the RCU should reserve a pool equal to the safety demand, which guarantees safety delivery at a quantified cost to M1. At the critical load, it should keep the pool shared, where per-vehicle learned placement does better than any partition. Choosing between the two needs the zone's current m0 and M --- information the RCU holds and no vehicle does.

\subsection{Result 6 --- Mixed Fleets (Mode 0b)}\label{sec:7.7}

Table~\ref{tab:4} reports the hybrid fleet at the critical load: two of the five safety vehicles and three of the five M1 vehicles are passive, each passive group served by one shared class actor, with the remaining vehicles running their own.

\begin{table*}[!t]
\centering
\caption{Mode 0b at $N = 10$ by vehicle type (three seeds; agent-specific critic). $^{*}$ interval clipped at 1.}
\label{tab:mode0b}
\label{tab:4}
\footnotesize
\begin{tabular}{ll}
\toprule
Vehicles & Delivery [95\% CI], or within-class collision by seed \\
\midrule
Safety, active (3 vehicles) & 0.941 [0.872, 1.000]$^{*}$ \\
Safety, passive (2, one shared actor) & 0.962 [0.926, 0.997] \\
M1, active (2 vehicles) & 0.964 [0.902, 1.000]$^{*}$ \\
M1, passive (3, one shared actor) & 0.825 [0.509, 1.000]$^{*}$ \\
\midrule
Passive safety pair, within-class collision & 0.004, 0.006, 0.002 \\
Passive M1 trio, within-class collision (random 0.360) & 0.372, 0.013, 0.026 \\
\midrule
Mode 0b fleet, safety & 0.9491 [0.9098, 0.9884] \\
Mode 0b fleet, M1 & 0.8803 [0.6617, 1.0000]$^{*}$ \\
All-Mode-0c fleet, safety (agent-specific) & 0.9413 [0.9113, 0.9713] \\
All-Mode-0c fleet, M1 (agent-specific) & 0.9645 [0.9576, 0.9715] \\
All-Mode-0a fleet, safety (agent-specific) & 0.6158 [0.6096, 0.6220] \\
\bottomrule
\end{tabular}
\end{table*}

The hybrid fleet keeps all five safety vehicles on separate subchannels in every seed (within-class 0.001) and delivers 0.9491 [0.9098, 0.9884] of safety messages --- a tie with the all-Mode-0c fleet on safety delivery, far above the all-Mode-0a fleet. On M1 delivery the hybrid's mean is lower, 0.8803 against 0.9645, but its interval is too wide to separate the two ([0.6617, 1.0000]*), because one seed's passive M1 trio stayed on its floor. Passive safety vehicles lose nothing to active ones: 0.962 against 0.941, with overlapping intervals, while the active vehicles match an all-Mode-0c fleet to within 0.001. The hybrid's costs fall elsewhere --- slower coordination (a median of 300 episodes against 50) and that seed's passive M1 trio (within-class 0.372; 0.013 and 0.026 in the others), the three-vehicle race of Section~\ref{sec:7.4}.

A passive vehicle is thus protected when its shared policy serves a small group: the passive safety pair separated in every seed. This gives the capability-tier-aware aggregation of Section~\ref{sec:5.2} a measured basis --- passive vehicles should be grouped finely, so that each shared policy serves few vehicles, while active vehicles resolve contention themselves. Protecting safety traffic in a Mode 0 zone does not require every vehicle to be upgraded.

\subsection{Result 7 --- Information Age}\label{sec:7.8}

\begin{figure*}[!t]
\centering
\includegraphics[width=\figwidew,keepaspectratio]%
                {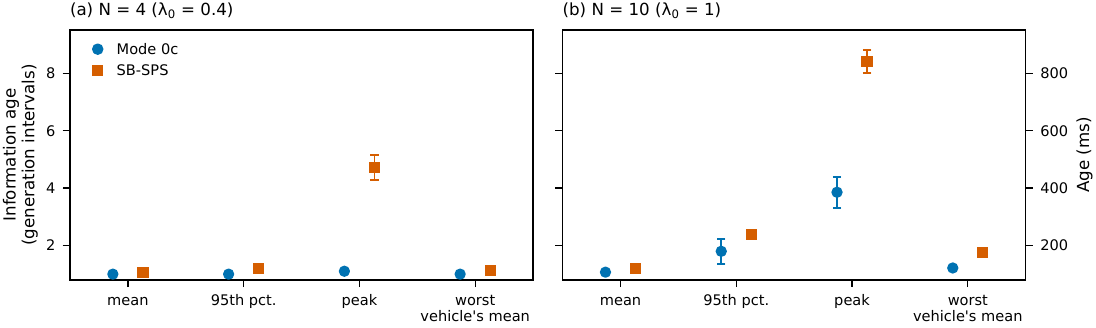}
\caption{Information age of safety messages at (a) $N = 4$ and (b) $N = 10$, Mode 0c against SB-SPS (five seeds each; 95\% intervals). Age is counted in generation intervals (right axis: ms), from delivery events drawn with each interval's delivery probability; 1 means refreshed in the current interval. The peak is each episode's longest staleness, averaged over episodes. Information age measures staleness, not latency.}
\label{fig:6}
\end{figure*}

Fig.~\ref{fig:6} reports how stale each vehicle's safety information becomes between deliveries.

At the critical load, the typical peak staleness is 3.85 [3.31, 4.40] generation intervals (message periods of 100 ms) under Mode 0c (about 385 ms) against 8.41 [8.01, 8.82] under SB-SPS (about 841 ms); the 95th percentile is 1.80 [1.36, 2.23] against 2.37 [2.28, 2.46], and the mean 1.069 against 1.204. Below the critical load the peak is 1.10 against 4.70. SB-SPS's losses therefore come in longer runs, which is the signature expected of semi-persistent reservations that keep colliding until reselection; this mechanism is inferred, not tested causally. Information age is not latency: the 20 ms requirement is addressed by the derivation of Section~\ref{sec:6.3}, and information age adds what consecutive losses do to the freshness of each vehicle's safety information.

\subsection{Discussion and Limitations}\label{sec:7.9}

The results support a narrower and sharper empirical case for Mode 0 than the architecture alone. Per-vehicle actors become necessary once a class group is too large for a shared policy to separate, which happens exactly at and above the critical safety load. Against the standardized baseline, their margin is largest in the tail --- at the worst interval, for the worst vehicle, and in peak staleness --- and it comes from coordinated subchannel choice rather than power control. The RCU's measured contribution lies on the execution side: a partition sized to the safety demand guarantees safety delivery where spectrum allows, which a weighted-sum learner does not, and deciding when to impose it needs zone-level knowledge. The critic's global view brings no measurable training advantage. Mixed fleets keep safety delivery for every vehicle.

Per-vehicle execution is not a computational obstacle: one actor decision costs 8.5 $\mu$s at the median and 16.9 $\mu$s at the 99th percentile on one CPU core in plain NumPy --- even the latter is 0.017\% of the 100 ms generation interval.

Several limitations bound these conclusions. First, the simulation tests scheduling only. It has no sensor occlusion geometry, so the two capabilities that depend on the RCU's elevated observation layer --- pre-emergence warning of occluded (``ghost peeking'') hazards and the trigger for Category C emergency escalation --- are argued architecturally and supported by deployment evidence (Sections~\ref{sec:3}--\ref{sec:5}), not evaluated here. Evaluating them by simulation, beginning with explicit occlusion geometry for the first, is follow-on work. Second, mobility is synthetic, and generalization is untested. Vehicles follow SUMO's default Krauss car-following and LC2013 lane-change models on a straight 3 km highway; no recorded trajectories or traffic traces were used, the scenario is not calibrated against traffic measurements, and classes are assigned in blocks by order of entry rather than interleaved (Section~\ref{sec:6.7}). Each run replays one traffic realization (Section~\ref{sec:6.7}), and the same $N$ vehicles remain agents throughout each episode, with no arrivals or departures. Training is intended to be node-specific --- an RCU's policies are trained for its own zone (Section~\ref{sec:8.7}) --- so transfer to other intersections is not something the architecture requires. Generalization within a zone is required, to traffic realizations, fleet mixes and densities the policies were not trained on, and it is not tested here. Channel shadowing and fading, by contrast, are drawn afresh in every interval, so every evaluation faces channel realizations the policies have not seen, though from a single channel model. Third, the radio model is abstracted. The resource grid is compressed to one opportunity per interval over five subchannels, so scarcity appears at ten vehicles; the comparison between schemes holds, but the density at which it applies does not transfer to NR's full grid. Propagation is line-of-sight only: the TR 37.885 state for links whose direct path is blocked by other vehicles (non-line-of-sight with vehicle blockage, NLOSv)~\cite{ref32} is not modeled, and the open highway has no building blockage; delivery is scored at a single receiver, the nearest vehicle, rather than at every vehicle in range; subchannels are orthogonal, so adjacent-channel interference is absent; there is no HARQ retransmission; half-duplex operation is represented only on a shared subchannel (Section~\ref{sec:6.4}) --- a vehicle can receive on one subchannel while transmitting on another, which an NR half-duplex radio cannot; the noise power is a fixed constant; and delivery follows a block-error curve whose parameters are assumed rather than fitted (Section~\ref{sec:6.4}). These choices set absolute delivery levels, and therefore the requirement verdicts and the reuse ratio, but they apply to every scheme alike. The arithmetic the conclusions rest on --- the random-choice floor, and whether a collision-free assignment exists --- does not depend on them, while the learned outcomes were observed under this abstraction only. A link abstraction calibrated against link-level simulation is follow-on work. Fourth, no signaling cost is modeled, although Mode 0 re-decides every interval --- the cost semi-persistence exists to avoid --- and compliance with RCU advice is assumed, the partition being enforced as a mask on each vehicle's choice. Fifth, coordination is budget-limited: separation at three vehicles per shared policy is a race that more training might win, and Mode 0c was not run beyond $N = 10$. Sixth, the reward-weight test covers one coefficient at one operating point, and the SB-SPS baseline is idealized as described in Section~\ref{sec:6.7}.

\textbf{Scope of the evaluation.} The division drawn in the third limitation applies to the environment as a whole. Absolute delivery, the spatial reuse ratio, the comparison with SB-SPS, and whether learning finds an available arrangement were observed in one environment; the arithmetic of fit was not. Urban intersections with occlusion would need a different road network, building geometry, a non-line-of-sight channel model, an RCU sensing model, and sub-zone partitioning. That is a separate simulation program, and the claims of this paper are scoped to match: scheduling is evaluated at the component level, while the information and authority dimensions are argued architecturally (Sections~\ref{sec:3}--\ref{sec:5}). We would expect occlusion to widen the advantage of Mode 0, since it degrades distributed sensing but not the view of an elevated RCU; this is an inference, not a result.

\section{Standardization Pathway and Open Issues}\label{sec:8}

\subsection{Where Mode 0 Fits in the 3GPP Work Program}\label{sec:8.1}

The proposed Mode 0 category does not require the construction of an entirely new standards track within 3GPP. The architectural components it relies upon --- PC5 sidelink communication, subchannel resource pools, UE-initiated resource selection, and infrastructure-assisted resource configuration --- are all already specified within the NR-V2X framework of Release 16 and its successors. What Mode 0 requires is the formal recognition of a third entity class alongside the base station and the UE, and the specification of the interface through which that entity exercises advisory scheduling authority. This is a standards extension, not a standards replacement, and its scope is narrower than the introduction of either NR-V2X Mode 1 or Mode 2, each of which required new physical layer specifications.

Mode 0 touches several working groups. The resource pool partition and the sidelink resource configuration belong in RAN1 and RAN2, where NR-V2X sidelink resource allocation is currently specified; the RCU as a new entity class and its interface to the operator's network belong in system architecture (SA2), and V2X application enablement in SA6. The advisory and escalation messages themselves are application-layer messages, like CAM, DENM and the SAE J2735 set: their formats would be defined in ETSI TC ITS and SAE in coordination with 3GPP, and 3GPP's part is to specify how they are carried and in which resources. A study item would address six standardization questions: (1) the formal definition of the RCU entity class and its three-layer functional decomposition; (2) the Mode 0 subfamily taxonomy (Mode 0a as the minimum compliance level, Mode 0b as the hybrid transitional category, Mode 0c as the advanced category) with capability declaration mechanisms by which UEs signal their individualized-policy capability; (3) the Mode 0 advisory message, its content and transport; (4) the emergency escalation protocol, including its message content, trigger conditions and the security of the escalation channel (Section~\ref{sec:8.8}); (5) the resource pool partitioning mechanism establishing the M0 communication resource subset, together with the cross-domain configuration interface, including the $M_{m_0}$ sizing interface, between the traffic management authority and the network operator (Sections~\ref{sec:8.2} and~\ref{sec:8.9}); and (6) the V2P extension (Section~\ref{sec:8.5}).

Mode 0a must be specified as the minimum compliance level, and the simulation shows what that requires. A deployment where all vehicle UEs are passive cannot rely on one shared learned policy serving many vehicles: at the critical load such a policy stays at the random-choice floor and delivers less than standardized sensing-based scheduling (Section~\ref{sec:7.2}). The minimum must therefore be specified through the grouping granularity of passive vehicles, so that each shared policy serves few of them, and through the safety partition where spectrum allows (Sections~\ref{sec:7.4},~\ref{sec:7.6} and ~\ref{sec:7.7}). These are the boundary conditions that all Mode 0 advisory and resource-partition specifications must satisfy. Mode 0c capability declaration allows the RCU to adapt its advisory recommendations to exploit per-vehicle actor coordination when the fleet composition permits it, without requiring any modification to the air interface specification.

\subsection{The Cross-Domain Interface Challenge}\label{sec:8.2}

The most significant standardization challenge that Mode 0 introduces is institutional rather than technical: the RCU sits at the boundary between telecommunications and transportation regulatory domains that have historically maintained separate standards ecosystems. Specifying Mode 0 requires defining an interface across this boundary.

On the telecommunications side, the cellular network operator manages the Uu interface and overall spectrum allocation, and needs to know which communication resources have been designated as the M0 resource pool for a given RCU zone. On the transportation side, the traffic management authority owns and operates the RCU, determines the M0 resource pool size, and manages the RCU's sensor infrastructure and advisory scheduling logic. Communicating this determination across these different regulatory frameworks currently has no standardized channel.

The co-deployment of traffic-authority RCUs and operator-owned RS-MEC units at the same traffic nodes --- documented in Section~\ref{sec:4.3} --- makes this cross-domain interface challenge concrete rather than hypothetical. At a Beijing intersection today, these two entities may share the same physical location with no standardized coordination protocol. Mode 0 provides an initial solution for this: the traffic authority configures the RCU's M0 resource pool, notifies the operator through a standardized management plane message, and the operator adjusts the RS-MEC's 5G/V2X resource allocation accordingly. The Chinese standards ecosystem documented in DB11/T 2329.1-2024~\cite{ref7} provides a useful model for the cross-domain data exchange protocol, with its explicit separation of RSU-to-cloud and MEC-to-cloud interfaces reflecting the institutional domain separation that Mode 0 formalizes at the resource allocation level.

Because the RCU's advisory scheduling authority is non-mandatory in its default form and does not modify vehicle transmission decisions, the cross-domain interface does not require the operator to cede control of any transmission resource. Mode 0 requires cross-domain coordination at the configuration level but preserves each domain's operational autonomy. The emergency escalation protocol of Section~\ref{sec:5.2} introduces the only exception: in Category C scenarios where mandatory advisory status is activated, the operator's RS-MEC must be informed that the M0 resource pool is under mandatory advisory mode, so that the operator's resource allocation decisions do not inadvertently compete with the mandatory advisory broadcast. This notification is a single status flag communicated through the existing cross-domain interface and does not require the operator to modify its transmission resources.

\subsection{Compliance Incentive Mechanisms}\label{sec:8.3}

Mode 0's baseline compliance model rests on individual rationality: a vehicle that follows the RCU's advisory receives better interference geometry than one that ignores it, because the advisory is derived from global zone information that no individual vehicle can replicate locally. This incentive does not require enforcement by any authority, although its strength grows with the share of vehicles that comply; it is an argument rather than a tested result (Section~\ref{sec:7.9}). The baseline individual rationality model is sufficient for the standards proposal.

\textbf{Compliance and fully autonomous vehicles.} The growing penetration of Level 4 and Level 5 autonomous vehicles (AVs) in urban traffic environments introduces an important asymmetry in the compliance landscape that Mode 0 deployment planning should address. A human driver receives a Mode 0 advisory and decides whether to follow it based on personal judgment, situational awareness, and risk tolerance --- the advisory is genuinely optional, a preference that can be implemented via a pre-set multi-level switch in the vehicle's system to dictate the degree of compliance with RCU advisories. An AV receives the same advisory and processes it algorithmically: by design of its V2X application layer, an AV can be configured to treat Mode 0 sub-zone recommendations as high-confidence inputs to its resource selection decisions, effectively making Mode 0 compliance near-certain without requiring any change to the advisory architecture's non-mandatory specification. It is important to clarify that this compliance advantage operates at the application layer and is independent of the Mode 0 subfamily architecture tier: an AV responding deterministically to Mode 0 sub-zone advisories is a high-quality Mode 0a-compliant participant --- it need not run per-vehicle MARL actors (Mode 0c) to achieve superior compliance. Mode 0 is therefore fully deployable in mixed autonomous/legacy fleets under Mode 0a, with Mode 0c providing additional coordination quality as fleet capability matures. As AV penetration increases, the aggregate compliance rate within a Mode 0 zone increases correspondingly, improving M0 interference coordination without any modification to the standard. In Category C emergency escalation scenarios, this property is particularly valuable: an AV fleet receiving a mandatory stop advisory responds with deterministic precision that a mixed human-AV fleet cannot guarantee.

Beyond this baseline, a complementary layer of economic incentives could accumulate across interactions to shape cooperative traffic behavior at the population level: an RCU-validated record of whether a vehicle follows advisories and reports hazards accurately could be redeemed through channels such as usage-based insurance (UBI) premium adjustment --- already commercially deployed at scale by major insurers --- and charging-fee deductions for electric vehicles, with distributed ledger technology as one plausible, non-normative implementation substrate for the tamper-resistant, cross-domain record-keeping such a scheme would require. Such a record is also privacy-relevant, since a contribution score tied to a vehicle identity can reveal information about a driver's behavior (see Section~\ref{sec:5.4} on data locality). Privacy-preserving reputation schemes address this problem. Examples are a dual reputation scheme for vehicle platoons that keeps reputation values, feedback scores and identities private~\cite{ref36}, and a scheme for updating reputations in cloud-assisted vehicular networks without exposing individual feedback~\cite{ref37}. Designing the privacy protection for a Mode 0 record is beyond the scope of this paper, but these works show the kind of mechanism it would need. This economic layer is identified as a research and deployment direction rather than a Mode 0 standardization requirement.

\subsection{Generational Roadmap}\label{sec:8.4}

The generational independence of Mode 0 --- established in Section~\ref{sec:5.3} --- has concrete implications for the standardization roadmap that we now develop.

Within the current 5G NR generation, Mode 0 would be standardized as an extension to the NR-V2X sidelink framework, using PC5 as the air interface for both RCU advisory broadcast and vehicle safety message transmission. The advisory message format, resource pool configuration protocol, emergency escalation message format, and cross-domain interface specification described in Sections~\ref{sec:5.1},~\ref{sec:5.2}, and 8.2 are all expressible within the NR framework without requiring new air interface work.

As 6G research and standardization progresses --- with 3GPP expected to begin formal 6G work items in the Release 21 timeframe --- Mode 0 as an architectural category does not need to be re-specified. The RCU entity class definition, the M0/M1 resource separation principle, the advisory scheduling and emergency escalation architecture, and the cross-domain interface structure are all defined at a level of abstraction independent of the specific air interface technology. What will require generational adaptation is the Mode 0 advisory message format and the specific resource pool configuration mechanism --- the equivalent of updating physical layer parameters when moving from LTE Mode 3 to NR Mode 1. The architectural category persists; the radio access technology instantiation is updated.

The generational independence of Mode 0 has an important practical implication for early deployment. Traffic management authorities considering RCU deployment do not need to wait for 6G standardization to invest in roadside computing infrastructure. An RCU deployed under the NR-V2X Mode 0 specification will continue to provide architectural value after the 6G transition, because its core functions --- including the Category C emergency escalation capability --- are not dependent on the specific air interface it communicates over. Durability applies to two of the three layers. The observation and evaluation layers --- sensors and computation --- do not depend on the air interface and persist across generations. The communication layer, the roadside unit's radio, would have to be replaced or upgraded at each generation, as base stations are. The durable part of the investment is therefore the sensing, the computation and the institutional framework, not the radio.

\subsection{V2P Extension and the Dual-Domain Pedestrian Problem}\label{sec:8.5}

The Mode 0 framework as specified in Section~\ref{sec:5} focuses on V2V safety scheduling as its primary scope. The extension of Mode 0 to V2P scenarios introduces a boundary condition that the current specification does not fully resolve and that represents one of the most consequential open problems for Mode 0 deployment.

\textbf{The detection distinction.} The first challenge is identifying, within the RCU's sensor field, whether a detected person is inside a vehicle --- and therefore fully represented by the vehicle UE for scheduling purposes --- or on the road as an independent traffic participant. The RCU's elevated sensor infrastructure can perform this distinction with reasonable accuracy in most scenarios: a detected human figure whose position does not correspond to any tracked vehicle position is classified as an on-road participant. However, edge cases introduce detection uncertainty: a pedestrian partially occluded by a stationary vehicle, a cyclist whose radar cross-section is ambiguous, or a person standing adjacent to their parked vehicle before crossing. These are detection accuracy requirements that must be specified as performance criteria for RCU sensor systems in future standardization work.

\textbf{The communication asymmetry.} Once correctly identified as an on-road participant, the pedestrian presents an architectural problem more fundamental than detection accuracy. Warning a vehicle UE about a hazard is straightforward within the Mode 0 framework: the RCU re-broadcasts the warning within the M0 PC5 sidelink resource pool, and the vehicle's OBU receives it over the same PC5 interface it uses for V2V communication. Warning a pedestrian requires a different communication path. A pedestrian's device --- typically a smartphone --- communicates primarily over the cellular Uu interface~\cite{ref38}. If the pedestrian's device supports PC5 sidelink (as specified for V2P scenarios in 3GPP Release 16 and later), direct PC5 communication from the RCU is possible. However, PC5-capable personal devices remain a small fraction of the installed device base, and relying on PC5 availability for pedestrian warning would leave the majority of pedestrians unreachable through the Mode 0 advisory channel.

The alternative --- routing the pedestrian warning through the eNB/gNB over the Uu interface --- reintroduces exactly the latency problem that Mode 0 is designed to avoid. A warning that travels from the RCU to the eNB/gNB and then to the pedestrian's device involves the multi-hop path whose worst-case latency variability was identified in Section~\ref{sec:4.1} as incompatible with 20 ms safety requirements.

\textbf{The dual-domain service object.} The pedestrian scenario reveals a deeper architectural tension in the Mode 0 framework. Section~\ref{sec:5.1} defines the RCU's service objects as entities participating in traffic behaviors within its zone. A pedestrian on the road unambiguously satisfies this definition. At the same time, the pedestrian is a subscriber to a network operator's Uu service, making them simultaneously a service object of the RCU (traffic authority domain) and a customer of the eNB/gNB (telecommunications domain). This dual membership is qualitatively different from the institutional boundary discussed in Section~\ref{sec:8.2}, which concerns infrastructure entities. Here the dual membership applies to an individual person.

Mode 0 is better positioned than either Mode 3/1 or Mode 4/2 to address V2P, because the RCU's sensor fusion can detect the pedestrian's position even when their device has no V2X capability whatsoever, and the RCU can issue a vehicle-directed warning based on sensor detection without any communication with the pedestrian's device. This is arguably the most important V2P safety function: warning vehicles about pedestrians, regardless of what device the pedestrian carries. The complementary function --- warning the pedestrian about approaching vehicles --- remains dependent on the pedestrian's device capability and the available communication path.

\textbf{Research directions.} Three directions are identified for future work. The first is a near-term practical mechanism for in-vehicle UE coverage that requires no change to personal device hardware: vehicle OBUs declare to the RCU the count of associated Uu subscriber devices within the vehicle while communicating over the PC5 interface. This allows the RCU to estimate in-vehicle occupant presence across its zone --- a conservative proxy, since one person often carries multiple devices, making the declared count an upper bound rather than an exact person count. Responsibility for accurate declaration is delegated to the vehicle OEM, whose V2X software stack is the correct architectural location for aggregating associated subscriber information. This mechanism addresses in-vehicle UE coverage; on-road pedestrian devices remain the open problem. The second direction is the specification of a lightweight RCU-to-pedestrian warning channel operating over Uu with latency optimization --- potentially using a reserved high-priority broadcast channel within the Mode 0 framework that the eNB/gNB serves with pre-committed resource allocation, extending the M0 resource separation principle to the Uu interface for pedestrian-directed safety messages. The third is the development of detection and classification algorithms for RCU sensor systems that reliably distinguish on-road participants from vehicle occupants, and the specification of minimum detection performance requirements as part of the RCU entity class definition.

The V2P extension is identified here as an open problem rather than a resolved component of the Mode 0 specification. A Mode 0 standardization study item should include the V2P extension as the sixth of its questions (Section~\ref{sec:8.1}), with the vehicle-directed pedestrian warning function as the near-term priority and the pedestrian-directed vehicle warning function as a medium-term objective dependent on the evolution of PC5-capable personal devices.

\subsection{Mode 0d and the Voluntary Downgrade Question}\label{sec:8.6}

The Mode 0 subfamily taxonomy defined in Section~\ref{sec:5.2} identifies three deployment tiers --- Mode 0a (all-passive), Mode 0b (hybrid), and Mode 0c (all-active) --- organized by the hardware capability of vehicle UEs to support per-vehicle actor operation. A fourth conceptual tier, tentatively designated Mode 0d, emerges from considering what happens when all vehicles in a zone are Mode 0c-capable but some voluntarily decline to exercise that capability in favor of a personalized driving preference. The Mode 0d scenario is: all-active UEs with partial voluntary downgrade.

This is a meaningful conceptual category and there is reason to anticipate that it will become practically relevant as fully connected and personalized autonomous vehicles proliferate. A vehicle owner might configure their vehicle's V2X participation mode based on privacy preferences, proprietary driving algorithm commitments, or commercial fleet management policies that conflict with transparent resource-selection reporting to the RCU. These are not hypothetical future concerns --- they are present today in the context of data-sharing policies for connected vehicles and are likely to intensify as the economic value of vehicle trajectory data grows.

What makes Mode 0d architecturally distinct from Mode 0b is not the resulting fleet capability composition --- both involve a mix of active and passive behaviors --- but the origin of the passivity. In Mode 0b, passivity is a hardware constraint: the vehicle cannot support per-vehicle actor operation. In Mode 0d, passivity is a software or policy choice: the vehicle can support it but opts out. This distinction matters for the RCU's advisory logic and for the standardization of capability declaration: a Mode 0d vehicle must accurately declare its elected participation mode rather than its hardware capability, and the RCU must be designed to handle a zone where declared capability and actual capability are decoupled.

The open questions that prevent Mode 0d from being formally specified at this stage are substantial. How does the RCU verify that a vehicle's participation mode declaration is honest, given that the vehicle has an incentive to under-declare capability if doing so reduces the demands placed on it? How does voluntary downgrade interact with the emergency escalation protocol of Section~\ref{sec:5.2}? A mandatory stop advisory asks for a driving response, which does not depend on a vehicle's resource-selection tier, so the question is whether a vehicle that has opted down to passive participation keeps the full obligation to respond to escalation. We would expect it to, since the declared tier governs only how the vehicle selects sidelink resources, but the escalation protocol would have to say so explicitly. And what is the appropriate regulatory treatment of voluntary downgrade in jurisdictions where V2X cooperation is mandated for road access?

These questions are flagged as future work for the Mode 0 standardization agenda. Mode 0d is a real and important category that the Mode 0 framework should eventually accommodate; the current paper does not attempt to resolve the open questions it raises, and Mode 0d is therefore excluded from the formal taxonomy of Section~\ref{sec:5.2} pending a more complete theoretical treatment in follow-on standardization work.

\subsection{Mobile Deployment and Compute Unit Scheduling}\label{sec:8.7}

The Mode 0 specification defines the RCU as a fixed infrastructure ensemble whose three layers are co-located at a single traffic node. This subsection examines a targeted relaxation of that assumption: mobility of the evaluation layer alone, and the constraints under which such mobility preserves the Mode 0 architectural guarantees.

\textbf{Compute-storage disaggregation and the node-bound model principle.} In distributed systems architecture, compute-storage disaggregation refers to the separation of processing resources from the data they operate on, allowing compute hardware to be dynamically allocated across locations while data remains in fixed storage. Applied to the RCU, this principle takes a specific and precise form. The policy model --- the actor network parameters, trained with the help of a critic on a specific node's intersection geometry, traffic flow patterns, sensor coverage envelope, and local channel conditions --- is node-specific in a deep sense: it encodes the characteristics of that particular traffic node and would not immediately generalize to a different intersection with different geometry or approach directions. This model is appropriately stored at the node's local persistent storage, not in the compute unit that executes it. The compute unit's role is to load the stored model, perform inference over the incoming sensor fusion stream, and generate advisory outputs --- functions that are computationally generic even though the model being executed is node-specific. A compute unit that departs the node carries no node-specific data with it; a new compute unit that arrives at the node loads the stored model and immediately assumes the full evaluation function. The model data is bound to the node. The hardware that executes it is not.

\textbf{Demand matching for geographically shifting traffic load.} The practical motivation for mobile evaluation layer deployment becomes clear when traffic demand is viewed over time rather than at a single instant. Predictable traffic tides --- holiday travel patterns, major event traffic, seasonal road condition variation --- create conditions where the total demand for RCU evaluation capability across a region is approximately constant at any given time, but the geographic distribution of that demand shifts substantially. The evaluation layer hardware required to serve peak demand at every node simultaneously would be prohibitively expensive and would be underutilized for the majority of the operational year. A pool of mobile evaluation units that can be relocated to high-demand nodes during peak periods --- and redistributed across the network during baseline conditions --- allows a smaller fleet of compute hardware to serve the aggregate regional demand without each node requiring permanently installed hardware sized for its individual peak load. The observation layer and communication layer at each node continue to operate during periods when no local compute unit is present: cameras and radar continue to capture environmental data, and the RSU continues to receive V2X messages from passing vehicles. The node stores this incoming data stream locally. When a compute unit arrives and loads the node's stored policy model, it processes not only the current sensor stream but also the buffered historical data from the period of absence --- allowing trajectory model updates and traffic pattern refinement that improve the advisory quality for the node's next active period. The mobile compute unit is therefore not merely a temporary substitute for permanent hardware; it is a mechanism for concentrating computational resources where demand is highest while maintaining observational continuity at all nodes regardless of compute unit presence. The deployment platform for mobile evaluation units is intentionally left unspecified at the Mode 0 standardization level, which defines only the functional interface between the evaluation layer and the observation and communication layers --- not the physical mechanism of transport. Purpose-built mobile platforms, vehicles equipped with standardized compute modules, or aerial platforms for rapidly relocating hardware to accident-affected intersections are all compatible with the same interface specification. This flexibility is consistent with the Mode 0 philosophy of defining roles and interfaces rather than prescribing implementation.

\textbf{Data privacy reinforcement through physical compute-data separation.} The mobile evaluation layer architecture reinforces the data locality guarantee established in Section~\ref{sec:5.4} by adding a second containment tier. Section~\ref{sec:5.4} established that the static RCU's zone boundary serves as a data governance boundary: vehicle kinematic data processed by the evaluation layer never leaves the zone under normal operating conditions. The mobile compute unit architecture adds node-level containment within the zone: the raw sensor observations and vehicle trajectory data that the evaluation layer processes are stored at the node and are never transferred to the compute unit's local memory beyond the duration of the active processing session. When the compute unit departs a node, it carries only updated model parameters --- the result of applying the stored policy to the observed data --- not the raw observations themselves. An adversary who intercepts the mobile compute unit in transit gains a device containing updated network weights and no identifiable vehicle trajectory data. This two-tier containment structure --- zone-level for all data, node-level for raw observations --- provides a privacy posture in which the raw data never leaves the node: the computation is performed, the advisory is generated, and the compute hardware is separated from the data before it leaves the node's physical custody. One qualification deserves acknowledgment: the updated model parameters that the mobile compute unit carries between sessions may encode aggregate statistical patterns about the traffic node's behavior --- the same gradient-leakage concern that motivates differential privacy techniques in federated learning. For deployment contexts where this residual risk is a concern, the Mode 0 framework is compatible with standard differential privacy mechanisms applied to parameter updates before the compute unit departs, which bound the amount of individual-level information that can be reconstructed from parameter inspection while preserving the advisory quality of the policy.

\textbf{Relationship to existing literature.} The mobile evaluation layer concept connects to an emerging body of work on mobile RSUs (mRSUs) and mobile edge computing nodes in the vehicular networking and ITS literature. Prior work on mobile secondary computing nodes (MSCNs) has explored dynamic allocation of compute resources across edge locations under MEC frameworks~\cite{ref4}. Separately, mobile RSU platforms --- vehicle-mounted or aerial --- have been proposed to extend communication coverage to underserved areas~\cite{ref39}. The novel contribution of the Mode 0 mobile evaluation layer is architecturally distinct from this prior work in one precise respect: existing mRSU proposals relocate the entire RSU function --- observation, communication and evaluation together --- as a unified mobile platform. The separation proposed here relocates only the evaluation layer while leaving the observation and communication layers permanently installed at the node. This separation is what makes the node-bound model principle possible: the observation and communication layers that generate the data remain fixed, ensuring that the stored policy model always reflects the specific conditions of the node at which it was trained, rather than the varying conditions encountered by a mobile platform traversing multiple locations. The mobility is in the hardware; the intelligence is in the node.

\subsection{Security of the Mandatory Escalation Channel}\label{sec:8.8}

The escalation protocol of Section~\ref{sec:5.2} gives the RCU mandatory authority under Category C conditions, which makes the escalation channel a high-value target. A forged mandatory stop would compel braking across several vehicles at highway speed and could cause the collisions the mechanism exists to prevent. Spoofing an advisory misleads a vehicle that can still judge for itself; spoofing an escalation compels it.

\textbf{Adversary and goals.} We adopt a Dolev-Yao adversary~\cite{ref40} that controls the wireless medium: it can eavesdrop on PC5 broadcasts and inject, replay, delay, or modify messages, but it cannot break the cryptographic primitives. We extend the model with compromise: the adversary may also obtain an RCU's credential, by theft or by compromising the RCU itself. The assets at risk are escalation messages, advisory messages, and the RCU credentials that authenticate both; a compromised credential is the most consequential, because it yields escalations that verify as authentic. An escalation message therefore needs four properties: authenticity (the receiver knows which RCU issued it), integrity (its content is unaltered), freshness (a captured message cannot be replayed later or elsewhere), and authorization (only an RCU entitled to escalate in a zone can issue an escalation for it).

\textbf{Mechanisms.} We do not design a new protocol; we map each goal to the existing V2X security ecosystem and state what must be added. Signatures under IEEE 1609.2~\cite{ref41}, with certificates issued through the Security Credential Management System (SCMS) in North America~\cite{ref42} and the ETSI trust and privacy management architecture~\cite{ref43} in Europe, give authenticity and integrity. Authorization rests on the certificate: in both ecosystems it states the application permissions it grants and can be limited to a validity period and a geographic region. Mode 0 would add an escalation permission --- a dedicated application identifier or service-specific permission that ordinary vehicle certificates do not carry --- granted only to an RCU entitled to escalate in its zone, in a certificate whose region is that zone. A receiver then accepts an escalation only under such a certificate, and only within its validity period and region. Freshness needs more than the signature, which still verifies when a message is replayed. IEEE 1609.2 lets a security profile require receivers to check a message's generation time, expiry time, and generation location and to detect replayed messages, but its messages carry no sequence number. The escalation profile must therefore make these checks mandatory, keep expiry times short, and add a sequence number per RCU, so that a receiver rejects an escalation generated outside the zone it is in, and one no newer than the latest message it has accepted from that RCU --- which also stops a replayed escalation from overriding a later status cancellation.

\textbf{Residual risks.} Authentication leaves two risks. First, a compromised RCU or a stolen credential issues escalations that verify as authentic. The defenses are detection by vehicles whose own sensing contradicts an escalation, a report to the misbehavior authority (a separate service in each ecosystem), and revocation. Vehicle-side detection is weakest exactly where escalation is needed, since a Category C hazard may lie beyond what vehicles can sense, and until revocation reaches the vehicles the credential is still accepted; short validity periods for escalation certificates bound that exposure independently of revocation. Second, jamming attacks the channel rather than the cryptography, and no authentication scheme prevents it: a jammed escalation is simply not received. Two questions also remain open. This paper gives no formal, machine-verified proof; the adversary model is the precondition for one, and verification with automated symbolic tools is future work. And whether revocation latency suffices for a channel whose failure compels behavior is unresolved; candidate extra layers are corroboration from more than one source before an RCU escalates, and a second, independent authorization for each escalation, for example from the traffic management authority. Authentication, replay protection, and formal verification of the escalation channel belong to the fourth question of Section~\ref{sec:8.1} and should be specified alongside the message definition in any Mode 0 standardization effort.

\subsection{The M0 Resource Pool Sizing Strategy and the Operator-Cooperation Precondition}\label{sec:8.9}

The sizing rule of Section~\ref{sec:7.6} states a target, not a procedure: it does not say how a traffic authority learns m0 and M in a live deployment, or when to switch between the safety partition and the shared pool. A concrete static procedure addresses the most immediate version of the question. The traffic authority sets $M_{m_0}$ from the node's own historical peak m0, observed over a representative monitoring period that spans typical weekday and weekend traffic, and revises it only when observed demand exceeds the current setting by a sustained margin across several monitoring periods. This hysteresis keeps the pool from reacting to transient fluctuations or single-day anomalies, at the cost of some lag after a persistent shift in demand. Section~\ref{sec:7.6} tests the sizing rule at three operating points but not this procedure, which sets $M_{m_0}$ from observed history; whether it tracks demand in a live deployment is untested.

Two questions remain open. Peak-based provisioning avoids the regime in which safety vehicles must share at all times, but it reserves spectrum that the operator's M1 pool cannot use during the much larger fraction of time when demand sits below peak, and at the critical load it would leave M1 no subchannel at all (Section~\ref{sec:7.6}). Dynamic resizing avoids that waste by tracking observed or predicted demand, but it raises its own questions: how often to resize, what triggers a resize, and how a change reaches the operator's M1 allocation without disrupting ongoing service. Spatial reuse (Section~\ref{sec:5.2}) may let a fixed $M_{m_0}$ absorb a range of vehicle counts without renegotiation, but the simulation models one contention pool rather than several sub-zones, so how far reuse reduces the need for resizing is untested (Section~\ref{sec:7.2}).

Sizing is also an institutional question. A traffic authority that provisions conservatively takes more M1 spectrum from the operator than necessary, which could erode the cooperation on which the cross-domain interface of Section~\ref{sec:8.2} depends. Resolving this needs either a resizing mechanism precise enough to avoid persistent over-provisioning or an arrangement in which the operator absorbs part of the cost, and this paper specifies neither. Dynamic resizing mechanisms and multi-RCU cross-domain coordination at scale are the subject of a companion specification currently in preparation, scoped to the operational ``how'' that this paper's architectural ``why'' and ``what'' leave open.

\subsection{Open Issues Beyond the Present Evaluation}\label{sec:8.10}

Several questions that the evaluation of this paper leaves open bear directly on the standardization agenda. The first is how the RCU estimates the zone's current safety demand and supply in real time and decides when to switch between a safety partition and a shared pool (Section~\ref{sec:8.9}). The second is the sensitivity of the learned policies to the reward weights where spectrum is spare, which the single-coefficient test of Section~\ref{sec:7.5} does not cover. The third is the coexistence of Mode 0 vehicles with legacy vehicles that run SB-SPS. The fourth is the evaluation of the sub-region hierarchy across several RCU zones. The fifth is the effect of occlusion and urban geometry on every result of Section~\ref{sec:7}, with road topology and mobility taken from recorded traffic traces rather than a synthetic highway (Section~\ref{sec:7.9}). The sixth is within-zone generalization --- to traffic realizations, fleet mixes, densities, and vehicles arriving and leaving --- together with the obligation that node-specific training creates: an RCU must be able to train or update its policies in operation, which simulated training time does not establish. The seventh is the signaling cost of per-interval recommendations, which the simulation does not charge. The eighth is a link abstraction calibrated against link-level simulation (Section~\ref{sec:7.9}).

\section{Conclusion}\label{sec:9}

The 3GPP V2X resource allocation framework, as currently specified across four modes spanning two cellular generations, rests on a binary entity taxonomy: the base station, which exercises centralized scheduling authority over vehicle UEs, and the vehicle UE, which exercises autonomous distributed scheduling without infrastructure involvement. This paper has demonstrated that this binary taxonomy is structurally incomplete. Neither entity class is architecturally capable of meeting the safety-critical V2X scheduling requirements that define the most demanding operational conditions --- the high-density traffic nodes where shared-resource contention degrades safety-class latency guarantees, the occluded hazard scenarios where road-level sensing cannot provide pre-emergence warning, the large-scope cascading environmental emergencies where advisory authority is insufficient and mandatory zone-wide coordination is required, and the mixed safety-and-entertainment traffic environments where M1 demand growth continuously erodes M0 service quality. The failure is not parametric; it cannot be resolved by optimizing schedulers within either existing entity class. It is categorical.

The missing entity is the Roadside Computing Unit, and the mode built around it is Mode 0, a new 3GPP V2X resource allocation category. The RCU is an infrastructure ensemble of three functionally necessary components: an elevated observation layer, a communication layer, and a local evaluation layer, operating as a unified system under traffic management authority ownership. It is neither a base station nor a UE. It is a third entity class whose authority is advisory by default, whose escalation to mandatory status is specified and principally justified for Category C large-scope hazard scenarios, whose optimization objective is traffic safety rather than network performance, and whose generational independence ensures that its architectural validity persists across the succession of cellular technology generations.

Mode 0 defines a family of deployment tiers --- Mode 0a (all-passive UEs, the minimum compliance level), Mode 0b (hybrid fleets, the transitional state), and Mode 0c (all-active UEs, the optimal target) --- enabling adoption across the full capability spectrum of current and future vehicle fleets. The bucket-effect argument governs the minimum compliance level: Mode 0a must function correctly even when the least-capable vehicle in the coverage zone determines the coordination floor, because in the 2026 traffic environment legacy vehicles without redundant onboard compute coexist with fully connected vehicles. It can do so only on a condition that the simulation identifies: the RCU must group passive vehicles finely, so that each shared policy serves few of them, or impose a safety partition where spectrum allows (Sections~\ref{sec:7.4},~\ref{sec:7.6} and ~\ref{sec:7.7}). Mode 0c represents the long-term aspiration, achievable as fleet modernization progresses.

Three independent lines of evidence support the Mode 0 proposal. Theoretically, the binary taxonomy fails under two distinct and complementary mechanisms: base station-led scheduling suffers resource contention saturation at traffic node densities, and Mode 4/2 UE autonomy is categorically incapable of providing pre-emergence warning for occluded hazards regardless of algorithmic sophistication. The traffic risk taxonomy --- small-scope contact risk (Categories A and B) and large-scope cascading environmental hazard (Category C) --- identifies the precise scenario class where advisory architecture is insufficient and where Mode 0's mandatory escalation capability is uniquely necessary --- the scenario class in which no existing 3GPP entity class can substitute for it. Empirically, Chinese local and industry-alliance standards including DB11/T 2329.1-2024 and T/ITS 0224.1-2025, alongside China Unicom's operator-owned RS-MEC deployment, European C-ITS programs, and US C-V2X initiatives, demonstrate that both sides of the institutional boundary are independently converging on the roadside traffic node --- traffic authorities deploying RCUs, operators deploying RS-MECs --- without a standardized coordination mechanism. Computationally, a multi-agent reinforcement learning simulation of the scheduling dimension (88 training runs, with 95\% confidence intervals over independent seeds) establishes four findings. Per-vehicle policies become necessary once a class group is too large for a shared policy to separate, which happens at and above the critical safety load. Against standardized sensing-based semi-persistent scheduling, their margin is largest in the tail --- at the worst interval, for the worst vehicle, and in peak information age --- and it comes from coordinated subchannel choice rather than power control. The measured contribution of the RCU lies in execution: a safety partition sized to the safety demand and imposed only where spectrum is spare delivers every safety message, while the global view of the critic brings no measurable advantage in training. Finally, mixed fleets keep safety delivery for passive and active vehicles alike.

The practical implication is direct. 3GPP should initiate a study item on Mode 0 as a new V2X resource allocation category within the NR-V2X sidelink enhancement work program. The study item's scope is the six questions of Section~\ref{sec:8.1}: entity class definition; Mode 0 subfamily specification (Mode 0a through Mode 0c, with UE capability declaration mechanisms); the advisory message; the emergency escalation protocol, including its message content, trigger conditions and channel security; the resource pool partition together with the cross-domain configuration interface, including the $M_{m_0}$ sizing interface between traffic management authorities and network operators; and the V2P extension. The standardization effort is bounded and tractable --- its components ground in existing sidelink specifications rather than requiring new air interface work. The need it addresses is not bounded: it grows with every vehicle that adds an infotainment screen, with every intersection that approaches peak density, with every Category C emergency that unfolds without a standards-compliant mechanism to halt it, and with every new RS-MEC and RCU deployment that proceeds without an interoperability framework. Mode 0 provides that framework.

\section*{Data and Code Availability}

The complete simulation program is openly available at https://github.com/Bluenight-jig/v2x-mode0-replication under the MIT license, and release v2.0.1, which contains the code and data behind every result in this paper, is archived at Zenodo under DOI 10.5281/zenodo.23242519 (all versions: 10.5281/zenodo.20338167). The release contains the SUMO scenario and the C++ channel-simulation bridge connected via ZeroMQ; the MAPPO trainers, including the per-agent advantage computation, the three critic variants, the Mode 0b hybrid-fleet trainer and the capability-ablation environments; the exact runner script behind each of the 200 training runs of the program, with a run register giving every run's role, results file and log; every run's evaluation results and training log; the SB-SPS baseline evaluations; the per-vehicle evaluation summaries and the trained policies they re-evaluate; and the scripts that compute every confidence interval and draw every figure and table in Section~\ref{sec:7}. From the archived release alone, these scripts regenerate every interval, figure and table reported here, as verified on a fresh clone. Of the 200 training runs in the register, 88 supply the results reported in Section~\ref{sec:7}, 11 more at the same final settings are archived as supplementary, and 101 were superseded by the corrections or used to diagnose them. Releases 1.x remain available but are superseded by release 2.0.1; the repository README explains why.

\balance

\end{document}